\documentclass[11pt,superscriptaddress]{article}

\usepackage[a4paper, total={5.8in, 9in}]{geometry}
\usepackage[indent,skip=4pt]{parskip}
\usepackage{amsmath,amssymb,amsfonts,amsthm,mathrsfs,mathtools,cases}
\usepackage{dsfont}
\usepackage{authblk}
\usepackage[colorlinks=true,linkcolor=blue, citecolor=blue, bookmarks]{hyperref}
\usepackage{cleveref}
\usepackage{microtype}
\usepackage{comment}
\usepackage{graphicx,adjustbox} 
\usepackage{subfig}
\usepackage{wrapfig}
\usepackage[font=small,labelfont=bf]{caption}
\usepackage[usenames,dvipsnames]{xcolor}  
\usepackage{booktabs}
\usepackage{braket}
\usepackage[nottoc]{tocbibind}
\usepackage{nicematrix}
\usepackage{tikz}
\usepackage{pgfplots}
\usepackage{cite}
\pgfplotsset{compat=1.18}
\usepackage{enumitem}

\usetikzlibrary{matrix}
\usetikzlibrary{decorations.markings}
\usetikzlibrary{patterns}
\usetikzlibrary{arrows.meta}
\usepackage{color,ifthen,amsthm,amsmath,amsxtra,amsfonts,dsfont,graphicx,bm,tikz,scalerel,wasysym,bbm,graphicx,amsthm,braket,physics,empheq,CircuitTikz,float}

\numberwithin{equation}{section}

\crefname{figure}{Fig.}{Figs.}
\crefname{equation}{Eq.}{Eqs.}
\crefname{section}{Sec.}{Secs.}
\crefname{appendix}{Appendix}{Appendices}
  
\colorlet{darkerblue}{blue!80!black}
\colorlet{lightblue}{blue!70!white}
\usepackage{braket}
\hypersetup{
    colorlinks,
    linkcolor={darkerblue},
    citecolor={darkerblue},
    urlcolor={darkerblue},
    backref=true
    }

\usepackage{CircuitTikz,braket}
\newcommand{\be}{\begin{equation}}
\newcommand{\ee}{\end{equation}}
\newcommand{\bea}{\begin{eqnarray}}
\newcommand{\eea}{\end{eqnarray}}
\newcommand{\bse}{\begin{subequations}}
\newcommand{\ese}{\end{subequations}}
\newcommand{\1}{\mathbbm{1}}
\usepackage{amsmath}
\newcommand{\sign}{\text{sign}}

\title{ Entanglement evolution from entangled multipodal states }
\author{Konstantinos Chalas$^1$, Pasquale Calabrese$^{1}$ and  Colin Rylands$^2$}

\date{}

\begin{document}
\maketitle
{\small
\vspace{-5mm}  \ \\
{$^{1}$}  SISSA and INFN Sezione di Trieste, via Bonomea 265, 34136 Trieste, Italy\\[0.1cm]
\medskip
{$^{2}$} Centre for Fluid and Complex Systems, Coventry University, Coventry, CV1 2TT, \\[-0.2cm]
\medskip \hspace{0.25 cm }United Kingdom\\[-0.1cm]
}

\maketitle

\begin{abstract}
In a periodic lattice system an entangled antipodal pair state, otherwise known as a crosscap state, is a simple two site product state in which spins at antipodal sites are prepared in Bell pairs. Such states have maximal bipartite entanglement and serve as a useful platform for studying the quench dynamics of systems which have large initial entanglement. In this paper, we study a generalization of these states which we dub entangled mutipodal states. These states, which are defined for fermionic systems, generalize the crosscap states by having correlations among more than two sites, specifically, those which sit at the vertices of regular polygons. By construction, the states are Gaussian and translationally invariant allowing many of their properties to be understood. We study the bipartite entanglement entropy of these states both in and out of equilibrium. In equilibrium, the entanglement profile as a function of subsystem size exhibits two distinct regimes, a volume-law growth followed by a saturation to a constant value, thus generalizing the Page-curve profile of the crosscap state. 
In the non-equilibrium setting, we study quenches from these initial states to the free-fermion chain, whose ensuing dynamics displays a far richer structure compared to the crosscap case. 
We interpret our results in terms of the quasiparticle picture, which requires multiplets of quasiparticles to be excited non-locally around the system. This scenario is confirmed by the appearance of a post-quench, negative tripartite information. 

\end{abstract}
\tableofcontents

\newpage
\section{Introduction}

Understanding the far-from-equilibrium dynamics of many body quantum matter has been, and still is, a significant endeavour for quantum physicists.
The most paradigmatic dynamical protocol to study such scenarios is the global quantum quench ~\cite{calabrese2006time,polkovnikov2011nonequilibirum}. There an initial pure state is unitarily evolved with a Hamiltonian, for which it is not an eigenstate. Usually, within this protocol, the initial states are chosen to be lowly entangled with the subsequent evolution observed to exhibit universal behaviour. In particular, they undergo  a linear growth in time of the bipartite entanglement entropy followed by a saturation to a volume law ~\cite{calabrese2016introduction,alba2017entanglement,alba2018entanglement,calabrese2020entanglement}.
This behaviour  has been seen to occur irrespective of the type of dynamics, e.g. in integrable or  chaotic systems with discrete time Floquet or continuous time Hamiltonian dynamics.
For free and interacting integrable systems, this behaviour is  elegantly captured by the quasiparticle picture of entanglement dynamics ~\cite{calabrese2005evolution,alba2017entanglement,alba2018entanglement} while for chaotic systems the entanglement membrane approach can be applied~\cite{jonay2018coarse,nahum2020entanglement}. More broadly, these features can be understood within the framework of space-time duality~\cite{bertini2022growth,bertini2023nonequilibrium,bertini2024dynamics}. 

Lowly entangled initial states are usually chosen because they can be treated efficiently both numerically, as well as analytically in some cases.
Some examples of such states are product states and ground states of local gapped Hamiltonians. Typically, these have a relatively simple structure and have been quite well understood with matrix product states introduced in tensor network techniques~\cite{cirac2021matrix}, which also allows one to simulate them efficiently, at least initially.
As mentioned above, when one considers quenches from such states, entanglement grows with time, leading to a more complicated description of the local state as time progresses. Even from a numerical perspective this growth presents a serious challenge~\cite{lerose2023overcoming,rath2023entanglement} as it can lead to exponential growth of the bond dimension, meaning the system is no longer simulable with classical means.
From this point of view,  initializing the system in a state with high amounts of entanglement appears to be incompatible with both numerical and analytical considerations.

Hence, on the one hand, it is not surprising that the quantum dynamics emerging from highly entangled states of matter has not been studied to the same extent as that from lowly entangled states, being limited to a small number of studies ~\cite{bucciantini2014quantum,kormos2014stationary, santalla2023entanglement,sierra_exotic_correlation_spread_2022,sierra2018_book_breakingarealawrainbow}.
On the other hand, states with large amounts of entanglement are realizable within currently available experimental platforms, such as trapped-ion systems~\cite{blatt2012quantum,monroe2021programmable,fossfeig2024progress} or Rydberg atom arrays, making experimental  realization of such quenches feasible.
Therefore, understanding dynamics of highly entangled initial states is a direction which can lead to novel dynamical behaviour in quantum many body systems, which can be experimentally realized, extending our knowledge of quantum dynamics beyond the study of lowly entangled initial states.

Moving in this direction and  despite the naive thought about quenches with highly entangled initial states being difficult to study, the quench dynamics from a class of highly entangled states called crosscap states has been initiated~\cite{chalas2024quenchdynamicsentanglementcrosscap,wei2025crosscapquenchesentanglementevolution,chen2025exact,bai2025spatially,dulac2025boundary}. In lattice systems with periodic boundary conditions, crosscap states or entangled antipodal pair states, as they are also called, are product states in which two sites of a  chain  are prepared in a Bell state. However, unlike typical scenarios, these sites are located at antipodal points in the chain, leading to maximal bipartite entanglement in the system. Despite their volume law  entanglement, the quench dynamics of such states can remain tractable using both analytical and numerical techniques ~\cite{chalas2024quenchdynamicsentanglementcrosscap,wei2025crosscapquenchesentanglementevolution,dulac2025boundary}. Previous studies have demonstrated that quenches to integrable models from these initial states can be effectively interpreted through a generalized quasiparticle picture, while in the case of maximally chaotic systems, the dynamics is accurately described by the entanglement membrane picture~\cite{chalas2024quenchdynamicsentanglementcrosscap} or holographic methods~\cite{wei2025crosscapquenchesentanglementevolution,dulac2025boundary}. 
See also a recent study~\cite{zhang2025entanglementgrowthentangledstates} which considers quenches from more general entangled initial  states.

In this paper, we wish to extend this program of quenches from highly entangled states to scenarios with richer dynamics while retaining a certain degree of solvability. We introduce a generalization of crosscap states, which we call {\textit{entangled multipodal states}}. These states are also defined on a periodic lattice system but instead of correlating sites which are antipodal, they couple sites which sit on the vertices of an $N$ sided polygon with the crosscap state being recovered at $N=2$. By changing $N$, the structure of the entanglement in these states can be tuned from the maximally entangled volume law starting at $N=2$, down to the area law for $N$ approaching the system size. For finite values of $N>2$ the states exhibit a saturation of entanglement as a function of subsystem size: while for small subsystem sizes the entanglement increases as in a volume law state, above a certain size it saturates to a value which depends upon $N$. Fermionic versions of these states are Gaussian which allows their quench dynamics to free fermion models to be studied exactly. For subsystem sizes which are smaller than $\frac{1}{N}$ of the total system size, the dynamics admits a modified quasiparticle picture form. For larger subsystem sizes, the dynamics has a marked difference, becoming more involved but can nevertheless be captured by a quasiparticle picture type expression. The nature of the expression in both regimes suggests the presence of entangled multiplets of quasiparticles rather than the more typical entangled pairs. This picture is confirmed by the existence of a post quench, negative, tripartite mutual information between symmetrically placed subsystems. 

The remainder of this paper is structured as follows. In Section~\ref{sec:crosscap} we review the notion of crosscap states in one-dimensional lattice models and the structure of their bipartite entanglement. In Section~\ref{sec:Polygon} we generalize these states and introduce the polygon crosscap states. We discuss their properties for fermionic systems including their entanglement and discuss spin chain realizations. In Section~\ref{sec:crosscap_dynamics} we discuss the quench protocol of interest and review some of the tools we will use in the context of the entanglement dynamics of crosscap states. In  section \ref{sec:multi_dynamics} we extend this analysis to the dynamics of the multipodal states and then provide a discussion in terms of the quasiparticle picture. In the final section, we conclude and comment on future directions. 

\section{Crosscap states}\label{sec:crosscap}

Our aim is to find novel states that exhibit large amounts of entanglement and lead to dynamics which can be analytically tractable.
This line of research builds on a recent and successful program developed for crosscap states in free-fermion quenches~\cite{chalas2024quenchdynamicsentanglementcrosscap,chen2025exact}, which demonstrated that even highly entangled initial conditions can lead to fully solvable time evolution in appropriate settings. Motivated by these results, we begin our analysis with a detailed review of crosscap states in equilibrium. This allows us to highlight the distinctive correlation structures they display and to examine how these features are encoded in both real-space and momentum-space representations of their fermionic counterparts.
 
\subsection{Spin crosscap states}
\begin{figure}
\centering
  {{ \includegraphics[width=0.95\linewidth]{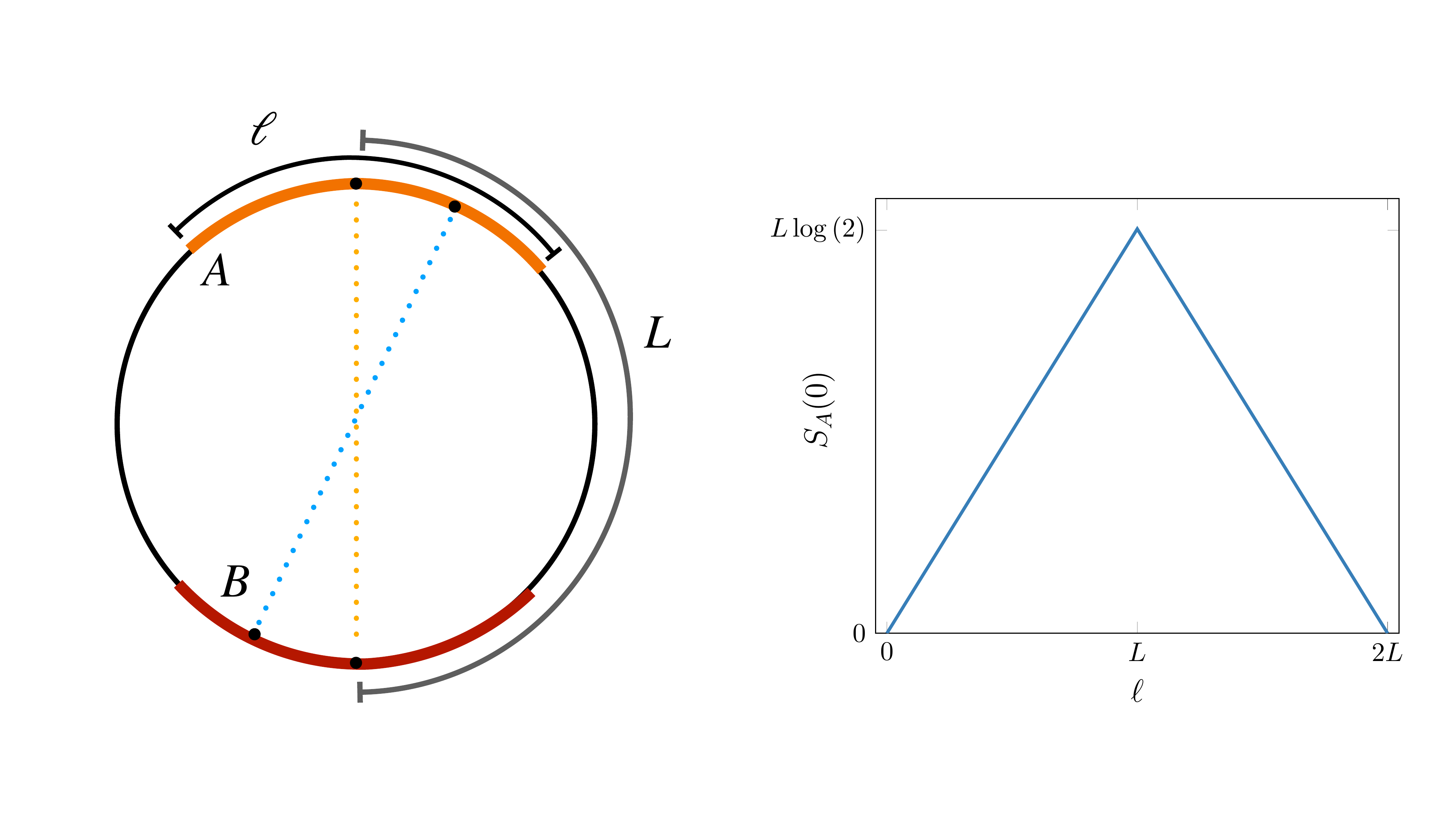} }}
    \caption{A depiction of the antipodal entangled or crosscap state. Antipodal sites, which are a distance $L$ apart are prepared in Bell pairs. Also depicted are the subsystem $A$ which is of length $\ell$ and its mirror $B$ which is diametrically opposite $A$. The structure of correlations generates maximal bipartite entanglement in the system as shown on the right hand side where we plot $S[\ket{\mathcal{B}_2}]$ as a function of $\ell$.}
    \label{fig:crosscap}
\end{figure}

Our system consists of a periodic lattice of $\mathcal{L}$ sites which for the case of the crosscap state is required to be even, e.g., $\mathcal{L}=NL$ with $N=2$. For our later generalizations, we shall require $N>2$.  The original lattice formulation of the crosscap state was in terms of a spin-$1/2$ chain and was defined as~\cite{caetano2022crosscap},
\begin{eqnarray}\label{eq:komatsu_integrable_crosscap}
\ket{\mathcal{C}_2}&=&2^{-L/2}\bigotimes_{x=1}^{L} \left(\ket{\downarrow}_x \otimes \ket{\downarrow}_{x+L}+\ket{\uparrow}_x \otimes \ket{\uparrow}_{x+L}\right)
    .
\end{eqnarray}
Here $\ket{\uparrow}_x,
\ket{\downarrow}_x$ are spin up and down states at site $x$. 
The form of these states, as well as their name, was inspired by work on CFTs in non-orientable manifolds, specifically those involving crosscap boundary conditions imposed at the ends of cylindrical manifolds
~\cite{fioravanti199sewing,tang2017universal,tang2018klein,Li2019critical,caetano2022crosscap}. This entails identifying antipodal points at the ends of the cylinders. The state above describes a discretized version of this. It is a chain of Bell pairs that are antipodally distributed on the chain in a translationally invariant manner, see Figure~\ref{fig:crosscap}. This can be generalized to allow for pairs to be prepared in other Bell states and also to larger local Hilbert space dimensions~\cite{chalas2024quenchdynamicsentanglementcrosscap}. The name entangled antipodal pair states has been introduced to cover these and other generalizations~\cite{yoneta_eap_PhysRevLett.133.170404,yoneta_eap_PhysRevResearch.6.L042062}. The above states were argued to be related to integrable boundary conditions for the Heisenberg spin chain~\cite{caetano2022crosscap} and therefore provide a set of integrable initial states for quenches in that model~\cite{piroli2017integrable}. Subsequently, the overlaps between $\ket{\mathcal{C}_2}$ and the eigenstates of the spin chain were exactly determined~\cite{gombor2022integrable1,ekman2022crosscap}, see also \cite{he2023integrable,gombor2022integrables2}. These states, as well as their deformations known as thermal pure states, have recently been shown to be exact eigenstates of certain spin chains in the mid-spectrum ~\cite{yoneta_eap_PhysRevLett.133.170404,yoneta_eap_PhysRevResearch.6.L042062,mestyán2025crosscapstatestunableentanglement}. Similar states also appear as quantum many-body scars in the PXP model ~\cite{ivanov_2025_scar_PRL}.

The entanglement structure of the crosscap states can be determined straightforwardly. For a subsystem $A$ of size $\ell\leq L$ the reduced density matrix is given by 
\begin{eqnarray}
    \rho_A={\rm tr}_{\bar{A}}\big[\ketbra{\mathcal{C}_2}{\mathcal{C}_2}\big]=\frac{1}{2^{\ell}}\1_A~.
\end{eqnarray}
Thus, the crosscap state is locally indistinguishable from the infinite temperature state.
This also implies that it has maximal bipartite entanglement entropy
\begin{equation}\label{eq:page_curve}
    S_{A}\big[\ket{\mathcal{C}_2}\big]=-\operatorname{tr}_A\big[\rho_A\log{\rho_A}\big]=\,{\rm min}[\ell,\mathcal{L}-\ell] \log{(2)}\,
\end{equation}
where we have used the reciprocity of bipartite entanglement, $S_A=S_{\bar{A}}$, to extend the result to all subsystems. This is the Page curve of entanglement entropy~\cite{page1993average} depicted in Figure \ref{fig:crosscap}. A straightforward way to probe the difference from the maximally mixed state is to compute the mutual information between a subsystem $A$ and its mirror, $B$, diametrically opposite to it, $\mathcal{I}_{A; B}=S_A+S_{B}-S_{A\cup B}$. Within this configuration, we have that 
\begin{eqnarray}
    \mathcal{I}_{A; B}\big[\ket{\mathcal{C}_2}\big]= {\rm min}[\ell,\mathcal{L}-\ell] 2\log(2)
\end{eqnarray}
which coincides with the maximum possible value of $\mathcal{I}_{A, B}$ and shows the existence of long range correlations between $A$ and $B$ as expected.

\subsection{Fermionic crosscap states}
It is convenient for the purposes of studying the quench dynamics of the crosscap state and its deformations to express it in terms of fermionic operators. Using the standard Jordan-Wigner transformation, one can show that $\ket{\mathcal{C}_2}$ is a sum of two fermionic Gaussian states related to each other via a gauge transformation~\cite{zhang2024crosscap}. Specifically, 
\begin{eqnarray}
    \ket{\mathcal{C}_2}=\frac{e^{i\pi/4}}{2^{(L+1)/2}}\prod_{x=1}^{L}\big(1+ic^{\dagger}_xc^{\dagger}_{x+L}\big)\ket{0}+\frac{e^{-i\pi/4}}{2^{(L+1)/2}}\prod_{x=1}^{L}\big(1-ic^{\dagger}_xc^{\dagger}_{x+L}\big)\ket{0}~,
\end{eqnarray}
 where $c^\dag_x, c_x$ are fermion creation and annihilation operators at site $x$ and $\ket{0}$ is the fermion vacuum, $c_x\ket{0}=0$. Taking just one of these states defines a fermionic crosscap state. We denote this by $\ket{\mathcal{B}_2}$ and after 
absorbing a phase in the definition of the fermions $c^\dagger_x\to e^{-i \pi/4}c^\dagger_x$, we have 
\begin{equation}
    \ket{\mathcal{B}_2}=2^{-L/2}\prod_{x=1}^{L}\big(1+c^{\dagger}_xc^{\dagger}_{x+L}\big)\ket{0}
    \label{eq:def_fermionic_crosscap}  ~. 
\end{equation}
We are interested in the properties of a  translationally invariant system, which, for the fermionic crosscap state written above, requires that the fermions obey anti periodic boundary conditions,
\begin{eqnarray}\label{eq:bc}
    c^\dagger_{x+\mathcal{L}}=-c^\dag_x,~~c_{x+\mathcal{L}}=-c_x~. 
\end{eqnarray}
  Using this, the fermionic crosscap state can be expressed as
\begin{eqnarray}\label{eq:fermionic_crosscap_def}
\ket{\mathcal{B}_2}&=&2^{-L/2}e^{\frac{1}{2}\sum_{x=1}^\mathcal{L}c^\dagger_x c^\dagger_{x+L}}\ket{0}\\
&=&\prod_{n=0}^{L-1}\frac{e^{\mathcal{B}_2(k_n)c^\dagger_{k_n} c^\dagger_{-k_n}}}{[1+|\mathcal{B}_2(k_n)|^2]^{\frac{1}{2}}}\ket{0}~,
\end{eqnarray}
where, in the first line we have extended the sum over the full system rather than half as in~\eqref{eq:def_fermionic_crosscap}. In the second line, we moved to Fourier space and introduced $c^\dagger_k=\frac{1}{\sqrt{\mathcal{L}}}\sum_{x=1}^{\mathcal{L}}e^{-ik x}c^\dagger_x$, the function $\mathcal{B}_2(k)=e^{ikL}$ and $k_n= 2\pi (n+1/2)/\mathcal{L}$ with $n=0,\dots, \mathcal{L}-1$.  The fermionic crosscap $\ket{\mathcal{B}_2}$ is therefore a squeezed state, highlighting its Gaussian nature. 

To obtain the entanglement entropy of the state for an arbitrary subsystem $A$ of length $\ell$, we exploit its Gaussianity and compute the two-point correlation matrix of the subsystem. Denoting this by  $C_{2}$, we find
\bea\label{eq:correlation_matrix}
[C_{2}]_{x,y}&=&\bra{\mathcal{B}_2} \textbf{c}^\dag_x\textbf{c}_y \ket{\mathcal{B}_2},\\
&=&\frac{1}{2} \begin{pmatrix}
     \delta_{x,y} & \delta_{|x-y|,L}\sign{(x-y)}\\
  \delta_{|x-y|,L}\sign{(y-x  )}& \delta_{x,y}  \end{pmatrix}
    \label{eq:corrmatr_crosscap},
\eea
where $\textbf{c}_x=(c_x,c^\dag_x)$ and we restrict $x,y\in A$. The above expression shows that there are correlations only on-site and between sites that are half of the total system size, $L$, apart.
Hence, $L$ is a characteristic length of correlations in the initial state. Denoting the eigenvalues of this matrix by $\lambda_\alpha, \alpha =1,\dots, 2\ell$ the entanglement entropy is given by~\cite{peschel2003calculation}
\begin{eqnarray}
    S_A=-\frac{1}{2}\sum_{\alpha=1}^{2\ell}\big[\lambda_\alpha \log \lambda_\alpha +(1-\lambda_\alpha)\log(1-\lambda_\alpha)\big]~.
\end{eqnarray}
By inspection, one can see that $C_2$ has $2\,{\rm min}[\ell,\mathcal{L}-\ell]$ eigenvalues equal to $1/2$ and $2\ell-2\,{\rm min}[\ell,\mathcal{L}-\ell]$ eigenvalues equal to zero. Accordingly, we once again have the entanglement of the Page curve~\eqref{eq:page_curve},
\be 
S_A\big[\ket{\mathcal{B}_2}\big]={\rm min}[\ell,\mathcal{L}-\ell] \log (2).
\ee 
The state lends itself easily to calculations of other quantities. For later comparison with our crosscap generalizations we look at the fermion mode occupation function,
\bea 
n(k_n)=\bra{\mathcal{B}_2}c^\dag_{k_n} c_{k_n}\ket{\mathcal{B}_2
}=\frac{|\mathcal{B}_2(k_n)|^2}{1+|\mathcal{B}_2(k_n)|^2}=\frac{1}{2}.
\eea
This is the same occupation function as the infinite temperature state even though the state only appears as such locally. The difference can be seen through the symmetry breaking which can  be probed through the anomalous correlator 
\begin{eqnarray}
    \Delta(k_n)=\bra{\mathcal{B}_2}c^\dag_{k_n} c^\dag_{-k_n}\ket{\mathcal{B}_2
}=\frac{\mathcal{B}^*_2(k_n)}{1+|\mathcal{B}_2(k_n)|^2}=\frac{1}{2},
\end{eqnarray}
which is distinct from the infinite temperature state.

\section{Entangled multipodal states}\label{sec:Polygon}
Having reviewed the notion of crosscap states in the previous section, we now introduce a geometric generalization of these, the entangled multipodal states. We do this using the fermionic form of the states, since those will be used when it comes to studying the quench dynamics.  We return to their formulation in terms of spins afterward.

\subsection{Fermionic  multipodal states}
 Previously, we saw that the fermionic crosscap state took the form of a Gaussian squeezed state on a lattice of size $\mathcal{L}=2L$, defined by the function $\mathcal{B}_2(k)$, see \eqref{eq:fermionic_crosscap_def}. We take this as our starting point and consider a lattice of $\mathcal{L}=NL$ sites, $N\in\mathbb{N}$. On this system, we denote the fermionic, entangled multipodal states by $\ket{\mathcal B_N}$ and define them in Fourier space to be \begin{eqnarray}\label{eq:polyogon_crosscap_def}
\ket{\mathcal{B}_N}&=&\prod_{n=0}^{{\lfloor\mathcal{L}/2\rfloor-1}}\frac{e^{\mathcal{B}_N(k_n)c^\dagger_{k_n} c^\dagger_{-k_n}}}{[1+|\mathcal{B}_N(k_n)|^2]^{\frac{1}{2}}}\ket{0},~~\mathcal{B}_N(k)=\sum_{m=1}^{N-1}e^{imkL}
~. \end{eqnarray}
Here we once again impose anti-periodic boundary conditions on the fermions~\eqref{eq:bc}. The anti commutation of the fermions  requires that, for consistency,  $\mathcal{B}_N(k)$ be an odd function $k$, $\mathcal{B}_N(-k_n)=-\mathcal{B}_N(k_n)$. This can be checked to hold due to the choice of boundary conditions which imposes $e^{i(N-m)k_nL}=-e^{im k_nL}$,  $\forall k_n$ and $m$. This definition naturally generalizes the crosscap definition and recovers~\eqref{eq:fermionic_crosscap_def} upon setting $N=2$. These states retain the properties of squeezed states and Gaussian states by construction and although the stated form obscures their real space structure, it can be easily discerned with the help of an example. 

\begin{figure}
\centering
  {{ \includegraphics[width=0.95\linewidth]{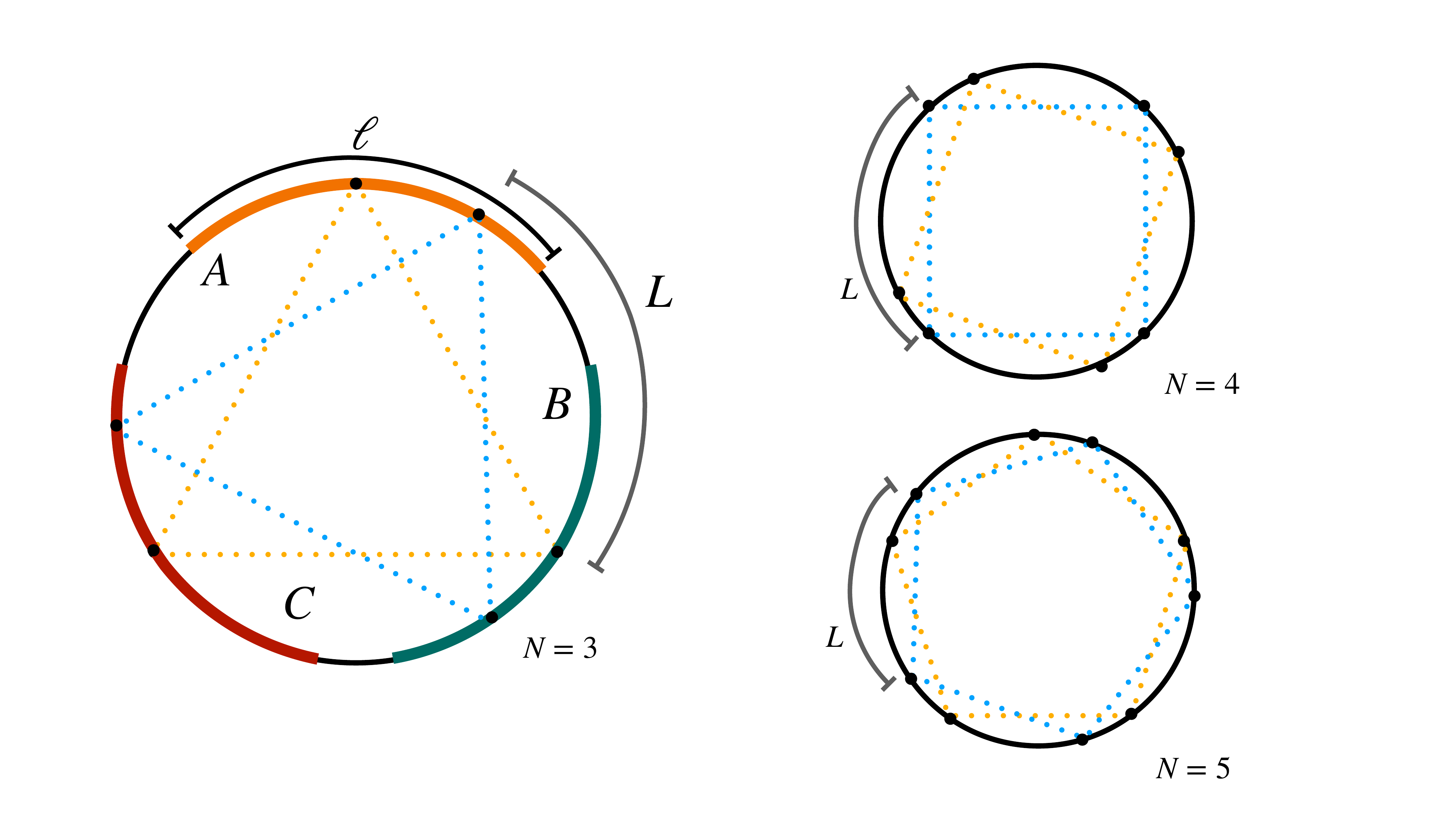} }}
    \caption{A depiction of the multipodal entangled states. On the left we have the triangle state, $N=3$ and on right we show the $N=4,5$ states. The multipodal entangled states exhibit correlations between sites which sit a distance $L$ apart, indicated by the dotted lines. On the left we mark the subsystems, $A,B,C$ which are all of equal length $\ell$. }
    \label{fig:triangle}
\end{figure}

 \textbf{Triangle state:} To understand the origin of the name, let us examine the first non-trivial member of this class of states beyond the crosscap state, $\ket{\mathcal B_3}$ which we call the triangle state. Explicitly, it is 
\begin{eqnarray}
    \ket{\mathcal{B}_3}&=&\prod_{n=0}^{{\lfloor\mathcal{L}/2\rfloor-1}}\frac{e^{ (e^{ik_nL}+e^{2ik_nL})c^\dagger_{k_n} c^\dagger_{-k_n}}}{[3+2 \cos(k_n L)]^{\frac{1}{2}}}\ket{0}\\
    &=& \frac{1}{4^{L/2}}\prod_{x=1}^{\mathcal{L}}\big(1+\frac{1}{3}\big[c^{\dagger}_{x}c^{\dagger}_{x+L}+c^{\dagger}_{x}c^{\dagger}_{x+2L}+c^{\dagger}_{x+L}c^{\dagger}_{x+2L}\big]\big)\ket{0}~.
\end{eqnarray}
From the real space expression, we can obtain the correlation matrix  of $\ket{\mathcal{B}_3}$ which we denote by $C_3$. We find 
\begin{equation}
    [C_3]_{x,y}= \frac{1}{4}\begin{pmatrix}
  2\delta_{x,y}+\delta_{|x-y|,L}-\delta_{|x-y|,2L}  & \sign{(x-y)}(\delta_{|x-y|,2L}-\delta_{|x-y|,L})  \\
 \sign{(y-x)}(\delta_{|x-y|,2L} -\delta_{|x-y|,L}) &  2\delta_{x,y}- \delta_{|x-y|,L}+ \delta_{|x-y|,2L} 
 \end{pmatrix},
 \label{eq:triangle_crosscap_eq_corrmatr}
\end{equation}
which has a structure similar to the crosscap state but with some important distinctions. Above we can note the existence of correlations between two distances $L$ and $2L$, corresponding to one third and two thirds of the system size respectively, as compared to  the crosscap state which has one length equal to half the system size. Thus, the correlated sites sit at the vertices of a equilateral triangle drawn on top of the periodic chain, justifying the name triangle state, see Figure~\ref{fig:triangle}.

\begin{figure}
\centering
  {{ \includegraphics[scale=0.99]{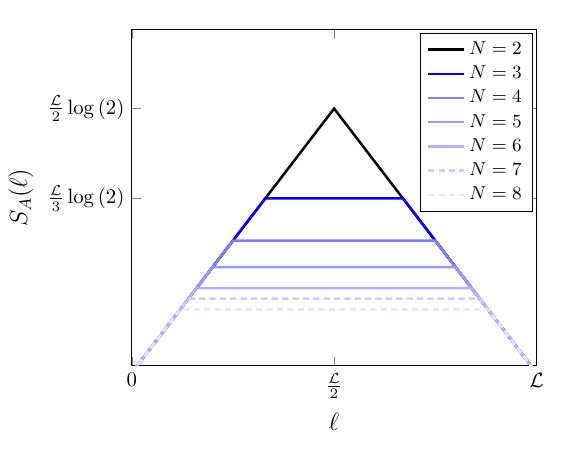} }}
    \caption{Entanglement entropy as a function of subsystem size for mutlipodal entangled states with  $N=2,3,4,5,6,7,8$.
    We notice that for each case the volume law saturates to a plateau at $\ell_A=L=\mathcal{L}/N$.}
    \label{fig:ee_polygon_crosscap_equilibrium}
\end{figure}

Since the state is Gaussian, the entanglement entropy can be found using the eigenvalues of $C_3$ as was done above. By inspection, one can check that $C_3$ has $2\,{\rm min}[\ell,\mathcal{L}/3,\mathcal{L}-\ell]$ eigenvalues equal to $1/2$ and $2\ell-2\,{\rm min}[\ell,\mathcal{L}/3,\mathcal{L}-\ell]$ eigenvalues which are zero. The resulting expression for the entanglement entropy is 
\begin{eqnarray}
    S_A\big[\ket{\mathcal{B}_3}\big]={\rm min}[\ell,\mathcal{L}/3,\mathcal{L}-\ell]\log(2)~.
\end{eqnarray}
Therefore, as a function of subsystem size, we see that the triangle state initially exhibits volume law entanglement but that this saturates at $\ell=L$. Following this, the entropy stays constant up to $\ell=2L$ when $|\bar A|=\ell$ and the entropy starts to decrease as expected due to the reciprocity of entanglement. This is depicted in Figure~\ref{fig:ee_polygon_crosscap_equilibrium}.  

We can also probe quantities beyond the bipartite entanglement entropy such as the bipartite and tripartite mutual information. For this we consider three intervals, $A,B,C$ of length $\ell<\mathcal{L}/3$ which are placed symmetrically about the system, see Figure~\ref{fig:triangle}. For the triangle state, we find that the bipartite mutual information between $A$ and $B$ is 
\begin{eqnarray}
    \mathcal{I}_{A;B}
    \big[\ket{\mathcal{B}_3}\big]=\ell \log(2)
\end{eqnarray}
while the tripartite mutual information defined as $\mathcal{I}_{A;B;C}\equiv\mathcal{I}_{A;B}+\mathcal{I}_{A;C}-\mathcal{I}_{A;B\cup C}$ is found to be
\begin{eqnarray}
    \mathcal{I}_{A;B;C}\big[\ket{\mathcal B_3}\big]=0.
\end{eqnarray}
As a result, we find that the triangle state has finite, but non maximal bipartite mutual information but vanishing tripartite correlations.

The squeezed state structure of the initial state once again allows us to calculate other quantities, in particular the occupation function and anomalous correlator. In this instance, the mode occupation function is
\begin{eqnarray}
    n(k_n)=\frac{|\mathcal{B}_3(k_n)|^2}{1+|\mathcal{B}_3(k_n)|^2}=\frac{2+2\cos (k_nL)}{3+2\cos (k_nL)}=\begin{cases}\frac{3}{4}~~&\text{if}~~n=0,2~\text{mod}\, 3\\
    0 ~~&\text{if} ~~n=1~\text{mod}\, 3
    \end{cases}.
\end{eqnarray}
while the anomalous correlator is
\begin{eqnarray}
    \Delta(k_n)=\frac{\mathcal{B}^*_3(k_n)}{1+|\mathcal{B}_3(k_n)|^2}=\frac{e^{-ik_nL}+e^{-2ik_nL}}{3+2\cos (k_nL)}=\begin{cases}-i\frac{\sqrt 3}{4}~~&\text{if}~~n=0~\text{mod}\, 3\\
    0 ~~&\text{if} ~~n=1~\text{mod}\, 3\\i\frac{\sqrt 3}{4}~~&\text{if}~~n=2 ~\text{mod}\, 3
    \end{cases}\label{eq:anomlaous_triangle}
\end{eqnarray}
where again $k_n=2\pi (n+1/2)/\mathcal{L}$. Rather remarkably, the occupation function in this state is discontinuous everywhere. However, despite this, the average occupation $\sum_{k_n}n(k)=\mathcal{L}/2$ which gives the correct expectation value of the particle number, which can be computed directly using the $\sum_{m=0}^{\mathcal{L}-1}n(k_m)=\sum_{x=1}^\mathcal{L}[C^{1,1}_3]_{x,x}$, where the summand is the top left component of ~\eqref{eq:triangle_crosscap_eq_corrmatr}. This observation along with the structure of the correlation matrix is reminiscent of ~\cite{trombettoni_gori_2015_volume_law_fermions}, where the authors find volume law free fermionic states which have similar properties. 

\textbf{General polygon:} The preceding analysis can be repeated for the general case with many of the same features emerging. For general $N$, the correlation matrix is
\begin{equation}
        [C_{N}]_{x,y}= \frac{1}{4}\begin{pmatrix}
  2\delta_{x,y}+\delta_{|x-y|,L}-\delta_{|x-y|,(N-1)L}  & \sign{(x-y)}(\delta_{|x-y|,(N-1)L}-\delta_{|x-y|,L})  \\
\!\sign{(y-x)}(\delta_{|x-y|,(N-1)L} - \delta_{|x-y|,L}) &  2\delta_{x,y}- \delta_{|x-y|,L}+ \delta_{|x-y|,(N-1)L} 
\end{pmatrix},
\label{eq:polygon_crosscap_correlation_equilibrium}
\end{equation}
where we observe that there are two correlation lengths present, $L$ and $(N-1)L$. We can  extend the geometric analogy we used in naming the triangle state: If we draw an $N$ sided polygon on the periodic chain, the state $\ket{\mathcal{B}_N}$ has correlations between a site which sits on a vertex and those which sit on the neighbouring vertices. Thus, every three consecutive vertices are correlated, like in the triangle state $\ket{\mathcal{B}_3}$, but now we have more such three-site-vertices, correlated in a translationally invariant manner, see Figure~\ref{fig:triangle}. Once again, the eigenvalue structure of $C_N$ gives us the entanglement entropy directly. The spectrum consists only of eigenvalues equal to $1/2$ or $0$, and the way their relative abundance evolves with subsystem size provides direct information on the scaling behaviour of the entanglement. Straightforwardly, we have 
\begin{eqnarray}
    S\big[\ket{\mathcal{B}_N}\big]={\rm min}[\ell,\mathcal{L}/N,\mathcal{L}-\ell]\log(2)
\end{eqnarray}
which is plotted for different values of $N$ in Figure~\ref{fig:ee_polygon_crosscap_equilibrium}. For finite $N$ this shows that for small enough subsystem size, the states have a volume law behaviour which then saturates for smaller system sizes as $N$ increases. In the limit $N\to \mathcal{L}$ the state becomes a short range entangled state. 

The occupation functions of these states are in general 
\begin{eqnarray}\label{eq:occup_func_polygon_crosscap}
n(k_n)&=&\frac{|\mathcal{B}_{N}(k_n)|^2}{1+|\mathcal{B}_{N}(k_n)|^2}=\frac{\sum_{l,m=1}^{N-1}e^{ik_nL(l-m)}}{\sum_{n,m=1}^{N-1}e^{ik_nL(l-m)}+1}~\\
\Delta(k_n)&=&\frac{\mathcal{B}^*_{N}(k)}{1+|\mathcal{B}_{N}(k)|^2}=\frac{\sum_{l=1}^{N-1}e^{-ilk_nL}}{\sum_{l,m=1}^{N-1}e^{ik_nL(l-m)}+1}
\end{eqnarray}
As was the case for the triangle states, this mode occupation function is nowhere continuous for finite $N$. Importantly, summing over all modes, we recover the correct average particle number $\sum_{k_n}n(k_n)=\mathcal{L}/2$. In the limit of $N\to \mathcal{L}\to\infty$, $n(k)$ approaches a smooth function.

We see that the multipodal states $\ket{\mathcal{B}_N}$, have a very intriguing structure which further generalizes the crosscap state, having two regimes for the behaviour of the  entanglement entropy as function of subsystem size.
This will lead to a difference in dynamical behaviour of the subsystem correlations depending on whether there was initially maximal entanglement or the subsystem was large enough to be outside this regime. 

\subsection{Multipodal spin states}

We have generalized the crosscap states by starting with their fermionic Gaussian form with a view to studying the entanglement dynamics. One can likewise examine spin versions of these states akin to the relationship between $\ket{\mathcal{B}_2}$ and $\ket{\mathcal{C}_2}$.  We focus on a single representative example, $N=3$. For the triangle state, the equivalent spin version can be defined as
\begin{eqnarray}  \nonumber\ket{\mathcal{C}_3}&=&4^{-L/2}\bigotimes_{x=1}^{L}\Big[\ket{\downarrow}_x\otimes\ket{\downarrow}_{x+L}\otimes\ket{\downarrow}_{x+2L}+\ket{\downarrow}_x\otimes\ket{\uparrow}_{x+L}\otimes\ket{\uparrow}_{x+2L}\\   &&\quad\quad\quad\,\,+\ket{\uparrow}_x\otimes\ket{\uparrow}_{x+L}\otimes\ket{\downarrow}_{x+2L}+\ket{\uparrow}_x\otimes\ket{\downarrow}_{x+L}\otimes\ket{\uparrow}_{x+2L}\Big]~.
    \label{eq:spin_triangle_crosscap}
\end{eqnarray}
This state features correlations between sets of three sites which are distributed around the chain in a translationally invariant manner. In such a state, the reduced density matrix can be found directly. For $\ell <\mathcal{L}/3$ we have 
\begin{eqnarray}
    \rho_A=\frac{1}{2^\ell}\1.
\end{eqnarray}
For large subsystems, $\rho_A$ is no longer diagonal, but the product state nature still allows one to compute it.  Upon doing so, we find the same entanglement profile as $\ket{\mathcal{B}_3}$ i.e $S_A\big[\ket{\mathcal{B}_3}\big]=S_A\big[\ket{\mathcal{C}_3}\big]$.  

We can also explore the mutual information in a straightforward manner by exploiting the product structure of the state, we again choose $\ell<\mathcal{L}/3$ and position the subsystems so that they are centred on the vertices a triangle. The result is the same as before, namely
\begin{eqnarray}
    \mathcal{I}_{A; B}\big[\ket{\mathcal{C}_3}\big]=\ell \log(2)~, ~\mathcal{I}_{A; B;C}\big[\ket{\mathcal{C}_3}\big]=0~.
\end{eqnarray}
Thus, the structure of these states is quite similar to the fermionic triangle state. 
Starting from $\ket{\mathcal{C}_2}$ and performing a Jordan Wigner transformation, we get two copies of the fermionic crosscap state. Likewise, we can take $\ket{\mathcal{C}_3}$ as a staring point and, after performing a Jordan Wigner transformation, see that the state cannot be written as a superposition of Gaussian fermionic states. This can be checked explicitly by considering the minimal example  of $L=2$.

\section{Entanglement dynamics of the crosscap state}\label{sec:crosscap_dynamics}

Our aim is to analyse the quench dynamics of the  multipodal states defined in the previous section.
Prior to this, we introduce the setup of our quench, the methods we use and briefly review the dynamics emerging from the crosscap state including its quasiparticle interpretation. 

\begin{figure}    
    \centering   
\centering   
    \centering   \includegraphics[width=0.49\linewidth]{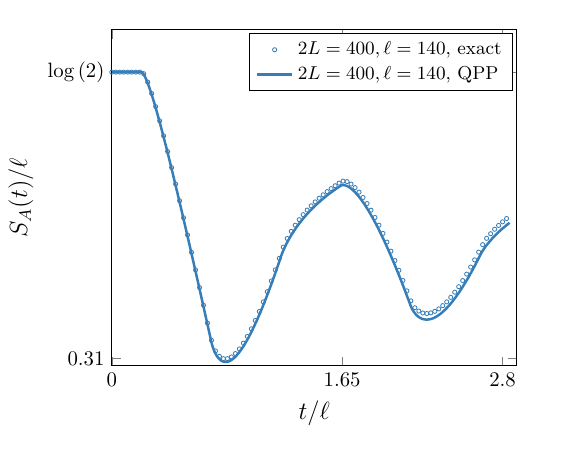}
    \includegraphics[width=0.49\linewidth]{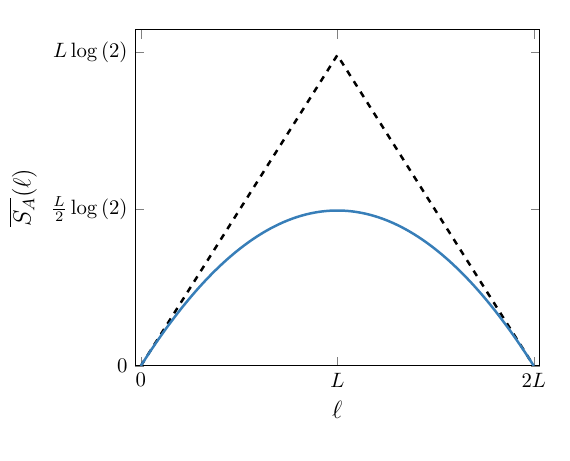}
    \caption{ On the left we show the entanglement dynamics from the crosscap state. The solid line is the prediction of the quasiparticle picture while the symbols are exact numerics. On the right we plot the long time average~\eqref{eq:crosscap_longtime}. The dashed line is the initial value for comparison.  }
\label{fig:EE_single_subsys_crosscap}
\end{figure}

\subsection{Hamiltonian and methods}
The dynamics of our system is governed by the free fermion tight binding Hamiltonian, 
\begin{eqnarray}\nonumber
 H&=&-\frac1{2}{}\sum_{x=1}^{\mathcal{L}} \left( c^\dagger_x c_{x+1}+c^\dagger_{x+1}c_x\right)\\&=&\sum_{n=0}^{\mathcal{L}-1}\epsilon(k_n)c^{\dagger}_{k_n}c_{k_n},
    \label{eq:diaged_tight_binding_apbc}
\end{eqnarray}
where, as above, the fermions obey anti-periodic boundary conditions and we have denoted the energy of a mode by $\epsilon(k_n)=\cos{(k_n)}$. We are interested in the quench dynamics emerging from the fermionic crosscap state $\ket{\mathcal{B}_2}$ which, since $H$ is quadratic and the state is Gaussian, can be fully understood from the time evolution of the two-point correlation matrix $C_2(t)$,
\begin{eqnarray}
    [C_2(t)]_{x,y}=\bra{\mathcal{B}_2(t)} \textbf{c}^\dag_x\textbf{c}_y \ket{\mathcal{B}_2(t)}
\end{eqnarray}
with $\ket{\mathcal{B}_2(t)}=e^{-iHt}\ket{\mathcal{B}_2}$. It will be useful to examine also the quantity $\Gamma_2(t)$ whose elements are
\begin{eqnarray}
  [\Gamma_2(t)]_{x,y}=2[C_2(t)]_{x,y}-\delta_{x,y}~.
\end{eqnarray}
A further simplification is that the crosscap state, along with all other states we consider, is a squeezed state, meaning it has the form 
\begin{equation}
\ket{\mathcal{M}}=\prod_{k>0}\frac{1+\mathcal M(k)c^{\dagger}_kc^{\dagger}_{-k}}{[1+|\mathcal{M}(k)|^2]^{\frac{1}{2}}}\ket{0},
    \label{eq:general_squeezed_state}
\end{equation}
where $\mathcal M(k)$ is an odd function of momentum, namely $\mathcal M(-k)=-\mathcal M(k)$ that defines the squeezed state. For such states, the subsystem correlation function can be worked out in general, giving
\begin{equation}
   [\Gamma(t)]_{x,y} =\frac{1}{\mathcal L}\sum_{m=0}^{\mathcal{L}-1} \frac{e^{-ik_m(x-y)}}{1+|\mathcal{M}(k)|^2} \begin{pmatrix}
   |\mathcal{M}(k)|^2-1  & 2{\mathcal{M}^{*}(k)}e^{-2i\epsilon(k_m)t}  \\
2 {\mathcal{M}(k) }e^{2i\epsilon(k_m)t} & 1- |\mathcal{M}(k)|^2
 \end{pmatrix},
 \label{eq:time_evolved_corrmatr_squeezed_state}
\end{equation}
where $x,y\in A$. As discussed in the previous section, the eigenvalues of this matrix can then be directly related to the entanglement entropy. 
Denoting the eigenvalues of $\Gamma(t)$ by $\gamma_{\alpha},\alpha=1,\dots, 2\ell$ we have that the entanglement entropy for a squeezed state is~\cite{peschel2003calculation}
\begin{eqnarray}\label{eq:ee_exact_peschel_eigenvals}
S_A\big[\ket{\mathcal M(t)}\big]\!\!&=&\!-\sum_{
\alpha=1}^{2\ell_A}\left[\frac{1-\gamma_\alpha(t)}{2}\log{\left(\frac{1-\gamma_\alpha(t)}{2}\right)}+\frac{1+\gamma_\alpha(t)}{2}\log{\left(\frac{1+\gamma_\alpha(t)}{2}\right)}\right]
    \,\,\\\label{eq:peschel_version2}
    &=&\!\ell\log(2)-\sum_{n=1}^{\infty}\frac{1}{2n(2n-1)}\operatorname{tr}[\Gamma(t)^{2n}]~.
\end{eqnarray}
The form presented in the first line is useful for obtaining the entropy via numerical methods. However, it is more  convenient when considering analytic methods to rewrite this as done in the second line. In particular, we shall be able to understand the dynamics of the entropy by considering the ratio of traces of powers of $\Gamma(t)$. 

\subsection{Quasiparticle picture}
Specializing~\eqref{eq:time_evolved_corrmatr_squeezed_state} to the case of the crosscap state using $\mathcal{M}(k)=e^{ikL}$, we find that the correlation function is 
 \begin{equation}
     [ \Gamma_2(t)]_{x,y} = \frac{1}{\mathcal{L}}\sum_{m=0}^{\mathcal{L}-1} e^{-ik_m(x-y)} \begin{pmatrix}
   0 & e^{ik_mL-2i\epsilon(k_m)t}  \\
  e^{-ik_mL+2i\epsilon(k_m)t} &0
 \end{pmatrix}.
 \label{eq:time_evolved_gamma_correlator_crosscap}
\end{equation}
In order to compute the entropy, we should  examine the quantities ${\rm tr}[\Gamma_2(t)^{2n}]$ which can be calculated in the scaling limit of $t,\ell\to \infty$ with $t/\ell$  and $L/\ell$ held fixed, using a multidimensional stationary phase approximation~\cite{fagotti2008evolution,fagotti2010entanglement}. The procedure requires some care  due to the presence of the highly oscillatory factors $e^{ik_mL}$. Nevertheless, it can be done and an analytic expression for the desired quantities can be found~\cite{chalas2024quenchdynamicsentanglementcrosscap}. A key aspect of the solution is seen already in the first term in the sequence, $ {\rm tr}[\Gamma_2(t)^{2}]$, for which we have
\begin{eqnarray}\label{eq:TrGamma2}
    {\rm tr}[\Gamma_2(t)^{2}]=\int_{-\pi}^{\pi} \frac{{\rm d}k}{2\pi}\left[  \chi_{L}(k,t)+ \chi_{-L}(k,t) \right],
\end{eqnarray}
where 
\begin{eqnarray}
    \chi_{z}(k,t)=\max{(0,\ell-|2v(k)\tau_{k}-z|)}
\end{eqnarray}
while $v_k=\partial_k \epsilon(k)=-\sin(k)$ is the mode velocity and $\tau_k=t~{\rm mod}{\frac{\mathcal{L}}{2v_k}}$. The functions $\chi_z(k,t)$ are known as counting functions. The appearance of terms like this is typical in quantities which obey a quasiparticle description and this form is typical for finite systems \cite{modak2020entanglement,Lagnese_2022}. We comment further on their meaning below, but prior to this, we can also note that the expression does not involve any $k$ dependence other than that which comes from $\chi_{\pm L}(k,t)$. This feature is also seen in the higher order terms which obey 
\begin{eqnarray}\label{eq:ratio_crosscap}
  {\rm tr}[\Gamma_2(t)^{2n+2}]={\rm tr}[\Gamma_2(t)^{2n}]~.
\end{eqnarray}
From this we determine that the eigenvalues of $\Gamma_2(t)$, $\gamma_\alpha$ can only take one of three possible values $0$ or $\pm 1$. Thus, the evolution only depends on how many of the $2\ell$ eigenvalues are zero as a function of time which is given by~\eqref{eq:TrGamma2}. Combining these elements, we have that the entanglement entropy in the scaling limit is 
\begin{equation}
        S_{A}\big[\ket{\mathcal{B}_2(t)}\big]=\ell\log{(2)}-2\log{(2)}\int_{-\pi}^{\pi} \frac{{\rm d}k}{2\pi}\left[ \chi_{L}(k,t)+ \chi_{-L}(k,t)  \right]~.
        \label{eq:crosscap_qpp}
\end{equation}
This expression was extensively checked against exact numerical evaluation of the entropy via~\eqref{eq:ee_exact_peschel_eigenvals} finding excellent agreement even for small system sizes~\cite{chalas2024quenchdynamicsentanglementcrosscap}, this is shown in Figure \ref{fig:EE_single_subsys_crosscap}. It admits a transparent physical interpretation in terms of the quasiparticle picture which was first elucidated in~\cite{chalas2024quenchdynamicsentanglementcrosscap} and can already be seen at the level of ${\rm tr}[\Gamma_2(t)^2]$. We will use a similar reasoning for $N>2$ and so we take some  time to briefly recap the arguments below. 

\begin{figure}
    \centering
    \includegraphics[width=0.8\linewidth]{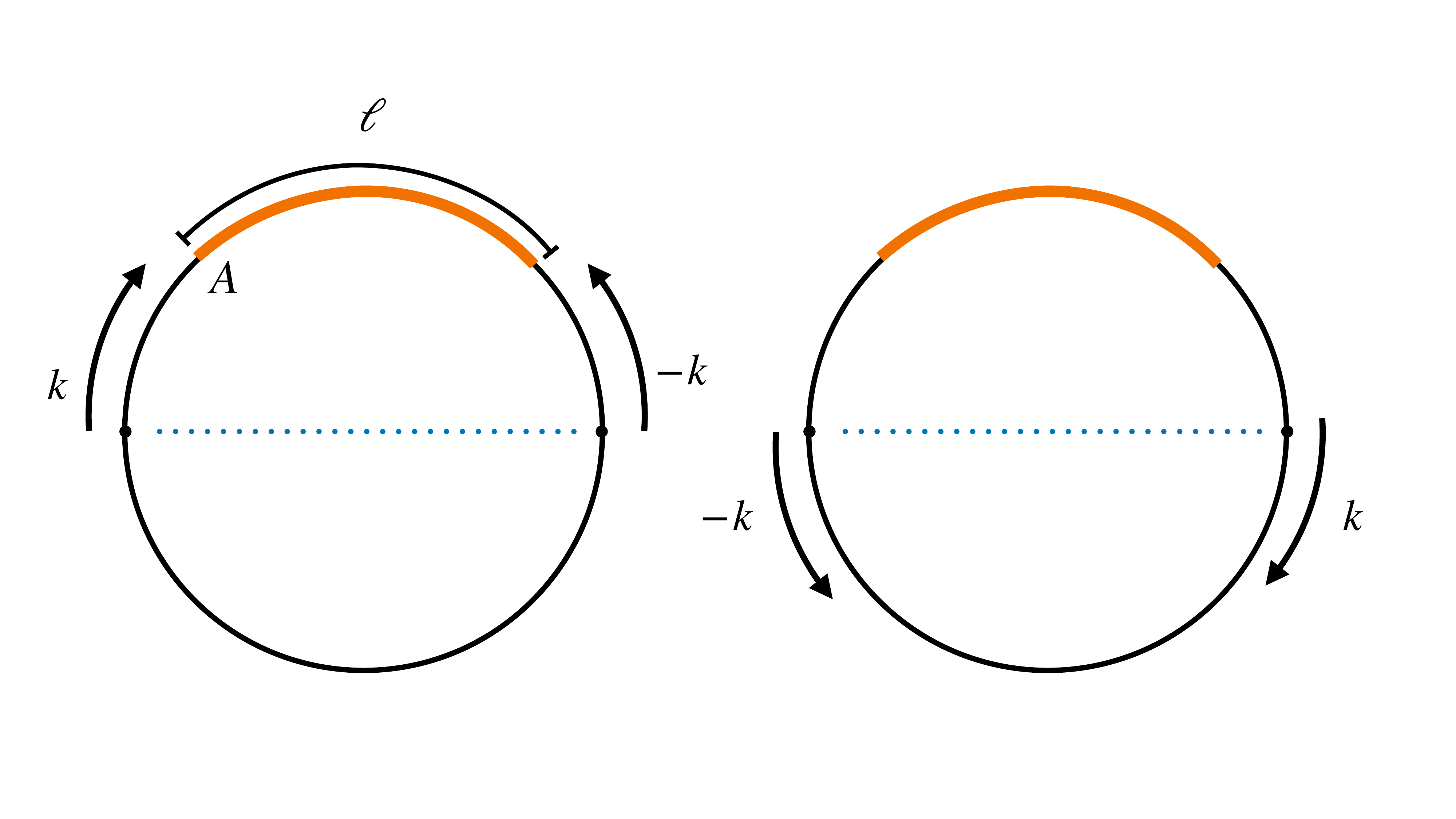}
    \caption{A depiction of the quasiparticle picture for the fermionic crosscap state quench. Entangled counter propagating quasiparticles  are created in pairs located at antipodes. The first change in the entanglement occurs when the pair (indicated by the arrows) on the left enters the subsystem. Subsequent drops in entropy occur when the pair emitted from the same point but travelling in the opposite direction, as shown on the right, enter the subsystem.  }
 \label{fig:QPP_crosscap}
    \end{figure}

The quasiparticle picture is a ballistic scale effective description which states that the quench excites stable quasiparticles which propagate in a semi-classical fashion throughout the system~\cite{calabrese2005evolution,alba2017entanglement,alba2018entanglement}. The correlations between quasiparticles are fixed by the initial state and various different quantities, including the reduced density matrix itself~\cite{Rottoli_2025,Travaglino_2025}, can be determined by knowing the initial correlations and then tracking how the quasiparticles are transported through the system. For the crosscap state, the quasiparticles are created in correlated pairs with members of a pair starting from antipodal sites on the circle and propagating toward each other, see Figure~\ref{fig:QPP_crosscap}. By contrast, a quench from a state with short range correlations, the correlated quasiparticles are emitted from the same point and travel in opposite directions. For more complicated short range initial state correlations, quasiparticles can be created in entangled multiplets, which, if are all emitted from the same point, are required to involve modes beyond $k,-k$~\cite{bertini2018entanglement,bastianello2018spreading,bastianello2020}. 

From the non-local nature of the initial state correlations, all the qualitative features of the entanglement dynamics shown in Figure~\ref{fig:EE_single_subsys_crosscap} can be understood. In particular, correlated quasiparticle pairs which are shared, i.e. one member in $A$ and the other in $\bar A$, generate entanglement between $A$ and $\bar A$ whereas pairs which are not shared do not. Initially, all pairs which are excited by the quench are shared since they are emitted from antipodal points. The entanglement can only change once pairs go from being shared to be not shared. The initial constant value seen in Figure~\ref{fig:EE_single_subsys_crosscap} appears because it takes a finite time for a pair to become sufficiently close so as to be completely contained within $A$ or $\bar A$. For a pair travelling with velocity $v_k$, the first time this can occur is $t^*=(L-\ell)/2 v_k$ and after this time the entanglement begins to decrease. Further features such as revivals can also be understood in this way. The functions $\chi_{z} (k,t)$ count the number of quasiparticle pairs which were initially separated by a distance $z$ and are completely  contained within $A$ at time $t$.  The periodic time variable ${\tau}_k$ accounts for both the finite size of the system, see Figure~\ref{fig:QPP_crosscap}. The drop in entanglement of $2\log(2)$ per pair can be understood from the fact that the pairs are maximally entangled and a single non shared pair replaces two shared ones inside the subsystem. This degree of entanglement between the pairs could be anticipated from the spectrum of $\Gamma_2$ whose eigenvalues switch from being $0$ to $1$ when pairs go from being shared to unshared.

\textbf{Long time average:} In order to witness the non-trivial behaviour of the entanglement, we need to explore times which scale with system size. As a result, and because the free fermionic system we are considering, the system does not equilibrate locally. Instead, we look at the long time average of the entanglement. From the quasiparticle expression one can find that 
\begin{eqnarray}
    \overline{S}_A[\ket{\mathcal{B}_2}]=\lim_{T\to\infty}\int_0^T\frac{{\rm d}t}{T} S_A\big[\ket{\mathcal{B}_2(t)}\big]=\log(2)\frac{\ell(\mathcal{L}-\ell)}{\mathcal{L}}\label{eq:crosscap_longtime}
\end{eqnarray}
where we have used 
\begin{eqnarray}\label{eq:time_aver_chi}
    \lim_{T\to\infty}\int_0^T\frac{{\rm d}t}{T}\int_{-\pi}^{\pi} \frac{{\rm d}k}{2\pi} \chi_{L}(k,t)=\frac{\ell^2}{4\mathcal{L}}~.
\end{eqnarray}
The long time average displays a characteristic parabolic shape as a function of $\ell$.  It is plotted on the right hand side of Figure~\ref{fig:EE_single_subsys_crosscap}.

\section{Entanglement dynamics of multipodal entangled states}\label{sec:multi_dynamics}
We now come to the main thrust of our work, the entanglement dynamics from multipodal entangled states. In Section~\ref{sec:Polygon} we observed that the equilibrium entanglement profile had a two-slope structure with initial maximally entangled volume law growth for $\ell<\mathcal{L}/N$ followed by a saturation to a constant value. We shall split the analysis of the dynamics accordingly into two sections, considering first initially maximally entangled subsystems and larger, non-maximally entangled, subsystems afterward. 

\subsection{Maximally entangled subsystem: $\ell<\mathcal{L}/N$}
We start by considering subsystems of size $\ell<\mathcal{L}/N$ so that $S[\ket{\mathcal B_N(0)}]=\ell \log(2)$ and look first at the triangle state $N=3$. To get the correlation matrix, we just need to substitute $\mathcal{B}_3(k)=e^{ikL}+e^{2ikL}$ into~\eqref{eq:time_evolved_corrmatr_squeezed_state} to get
\begin{equation}
   [\Gamma_3(t)]_{x,y}=\!\sum_{m=0}^{\mathcal{L}-1} \frac{e^{-ik_m(x-y)}/L}{1+4\sin^2\!\big(\frac{k_mL}{2}\big)} \begin{pmatrix}
   2\sin^2\!\big(\frac{k_mL}{2}\big)-\frac{1}{2} \!\!\! & \!\!\!\!\!\!\!2(e^{ik_mL}+e^{2ik_mL})e^{-2i\epsilon(k_m)t}  \\
2 (e^{-ik_mL}+e^{-2ik_mL}) e^{2i\epsilon(k_m)t} & \frac{1}{2}-2\sin^2\!\big(\frac{k_mL}{2}\big)
 \end{pmatrix}\!.
 \label{eq:time_evolved_corrmatr_triangle_crosscap_state}
\end{equation}
We shall use the multidimensional stationary phase approximation to evaluate tr$[\Gamma_3(t)^{2}]$ in the limit $t,\ell\to \infty$ holding $t/\ell$ fixed. The calculation follows that of the crosscap state, but extra care must be taken to properly account for the rapidly oscillating terms which appear in the diagonal components of~\eqref{eq:time_evolved_corrmatr_triangle_crosscap_state}. The main feature one can observe from the result is that in the ballistic scaling limit and within the regime of validity of the stationary phase approximation
\begin{eqnarray}\label{eq:triangle_Gamm_2}
     {\rm tr}[\Gamma_3(t)^{2}]=\int_{-\pi}^{\pi} \frac{{\rm d}k}{4\pi}\left[  \chi_{L}(k,t)+ \chi_{2L}(k,t)+ \chi_{-L}(k,t)+ \chi_{-2L}(k,t) \right]~.
\end{eqnarray}
From this we highlight some features. First, we see the appearance of the quasiparticle occupation functions $\chi_{\pm L}(k,t)$, which were present for the crosscap state, and in addition we also have  $\chi_{\pm 2L}(k,t)$ which are new. Moreover, we see that, as with the crosscap state, the integral has no $k$ dependence other than that of the counting functions. Going beyond the lowest order to calculate ${\rm tr}[\Gamma_3(t)^{2}]$ is complicated analytically but numerically can be straightforwardly carried out. From this one observes that   
\begin{eqnarray}
   {\rm tr}[\Gamma_3(t)^{2n+2}]\simeq\frac{1}{4} {\rm tr}[\Gamma_3(t)^{2n}]~,
\end{eqnarray}
which can be compared with the result for the crosscap state~\eqref{eq:ratio_crosscap} where all the traces were the same. From this we determine that all the eigenvalues of $\Gamma_3(t)$ are either $0$ or $\pm\frac{1}{2}$, with the dynamics coming from the relative number of zero to non-zero eigenvalues which is given by~\eqref{eq:triangle_Gamm_2}. 

Inserting this into~\eqref{eq:peschel_version2} we arrive at the expression for the entanglement entropy
\begin{equation}
    S_A\big[\ket{\mathcal B_3(t)}\big]=\ell\log{(2)}-\log{\left(\frac{27}{16}\right)}\int_{-\pi}^{\pi}\frac{{\rm d}k}{8\pi}\left(\chi_{L}(k)+\chi_{2L}(k)+\chi_{-L}(k)+\chi_{-2L}(k)\right)
    \label{eq:qpp_triangle}
\end{equation}
Contrasting with~\eqref{eq:crosscap_qpp} we immediately  notice several modifications. First we have the appearance of counting functions at additional length scales $\chi_{z}(k,t)$  with $z=\pm L,\pm 2L$. These are inherited from the behaviour of ${\rm tr}[\Gamma_2(t)^2]$ and directly follows from the additional length scale which appears in the off-diagonal elements of  the initial state correlation matrix~\eqref{eq:correlation_matrix}.  The second difference is the unusual seeming numerical factor $(1/4)\log (27/16)$ which weights the counting functions. This can be traced back to the spectrum of the correlation matrix $\Gamma_3$ acquiring eigenvalues $\pm 1/2$ through the course of the evolution. 

\begin{figure}    
    \centering   
\centering   
    \centering   \includegraphics[width=0.49\linewidth]{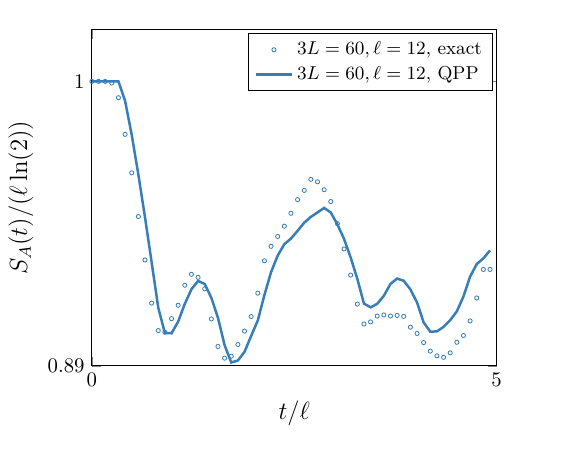}
    \includegraphics[width=0.49\linewidth]{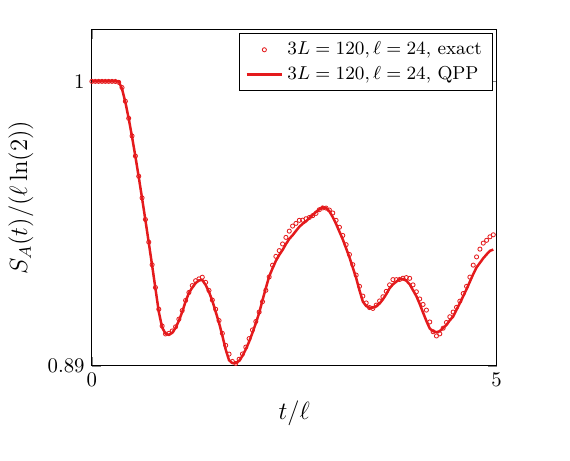}
    \includegraphics[width=0.49\linewidth]{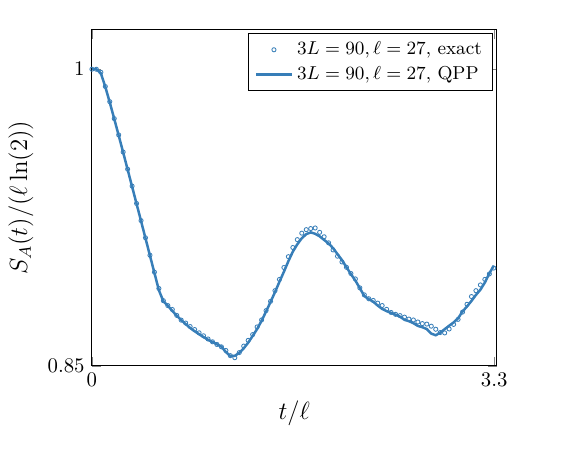}
    \includegraphics[width=0.49\linewidth]{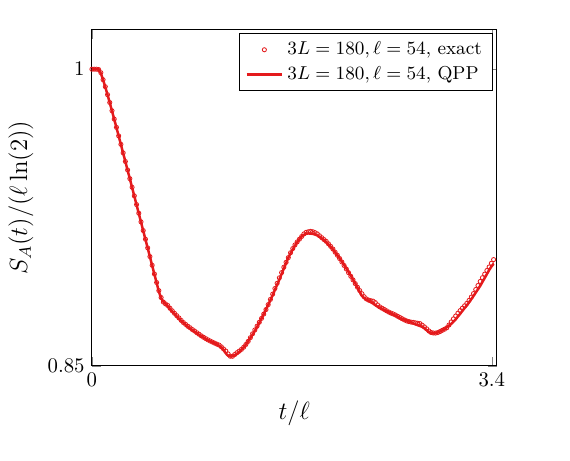}
    \caption{Triangle state entanglement dynamics ($N=3)$ for $\ell<L$ (initially maximally entangled subsystem) for different system and subsystem sizes. We  see that the prediction~\ref{eq:qpp_triangle} prediction (solid lines) converges to the exact results (symbols) for increasing subsystem and system size. Here we have the ratio $\frac{\ell}{\mathcal{L}}=0.2$ (top row) and  $\frac{\ell}{\mathcal{L}}=0.3$ (bottom row) with the agreement improving as we approach $\ell/\mathcal{L}=1/3$. }
\label{fig:EE_single_subsys_beta1_triangle_ell_less_L}
\end{figure}

In Figure~\ref{fig:EE_single_subsys_beta1_triangle_ell_less_L} we plot $S_A[\ket{\mathcal{B}_3(t)}]$ for different system sizes comparing the exact numerical result with the prediction of~\eqref{eq:qpp_triangle}. We see that the agreement between the two becomes very good as we increase the subsystem and system sizes and also improves as we approach $\ell/\mathcal{L}=1/3$. 
We shall comment on the quasiparticle interpretation of the expression~\eqref{eq:qpp_triangle} below, but now proceed to look at the more general states.

\begin{figure}    
    \centering   
\centering   
    \centering   \includegraphics[width=0.49\linewidth]{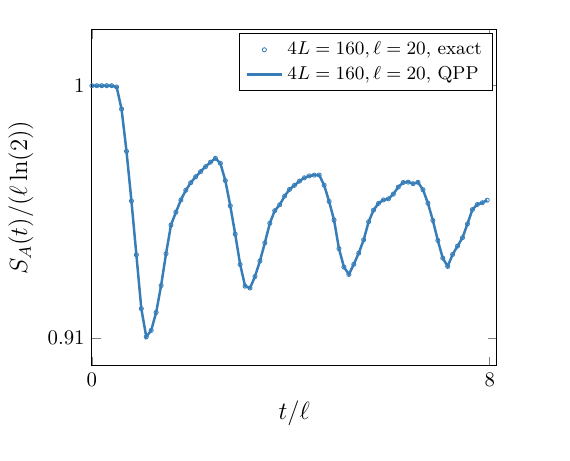}
    \includegraphics[width=0.49\linewidth]{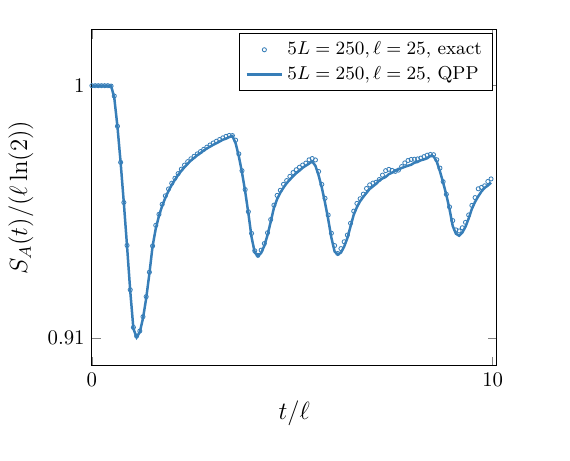}
    \includegraphics[width=0.49\linewidth]{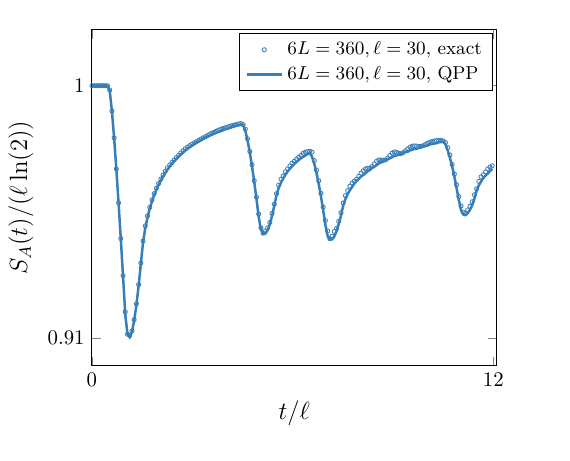}    
    \includegraphics[width=0.49\linewidth]{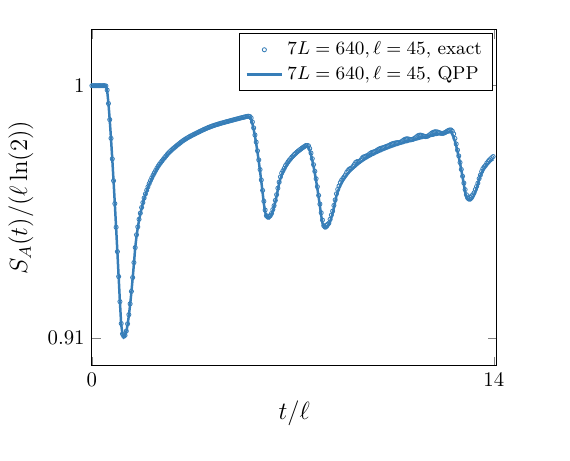}  
            
    \caption{Entanglement dynamics from multipodal entangled states for $\ell<L$ with $N=4,5,6,7$ clockwise from the top left. The symbols are exact numerics while the solid lines are the prediction of~\eqref{eq:qpp_polygon}}. 
\label{fig:EE_single_subsys_polygonsbeta_ell_less_L}
\end{figure}

\textbf{General multipodal state dynamics:} 
We have seen that aspects of dynamics of the entanglement entropy in the triangle state can be traced back quite simply to the structure of the correlation matrix. In particular, the presence of the two length scales, $L$ and  $2L$, in~\eqref{eq:triangle_crosscap_eq_corrmatr} gave rise to the two types of counting functions in the expression~\eqref{eq:qpp_triangle} while the appearance of eigenvalues of $\pm 1/2$ in $\Gamma_3(t)$ leads to the unusual weighting. Given that the more general polygon states exhibit the same structures, it is quite natural to expect that the dynamics be broadly similar. Indeed, the expression~\eqref{eq:triangle_Gamm_2} holds for arbitrary $N>2$ and one can also check numerically that the ratio formula applies to general $N$ as well,
    \begin{eqnarray}
  {\rm tr}[\Gamma_N(t)^{2n+2}] \simeq\frac{1}{4} {\rm tr}[\Gamma_N(t)^{2n}]~.
\end{eqnarray}
Combining these elements, we find that for a general multipodal state $N>2$ we have 
\begin{equation}
    S_A\big[\ket{\mathcal B_N(t)}\big]=\ell\log{(2)}-\log{\left(\frac{27}{16}\right)}\int_{-\pi}^{\pi}\frac{{\rm d}k}{8\pi}\left(\chi_{L}(k)+\chi_{\mathcal{L}-L}(k)+\chi_{-L}(k)+\chi_{L-\mathcal{L}}(k)\right)
    \label{eq:qpp_polygon}
\end{equation}
which is more or less the same as the triangle expression but with counting functions with length scales $\pm L$ and $\pm(\mathcal{L}-L)=\pm(N-1)L$. This expression is valid for any $N>2$,  reproducing the triangle expression at $N=3$. In Figure~\ref{fig:EE_single_subsys_polygonsbeta_ell_less_L},  we plot the entanglement as function of time for several different states $N=4,5,6,7$. We compare the results of~\eqref{eq:qpp_polygon} with those obtained using exact numerical methods for different system sizes, showing excellent agreement.

\subsection{Non-maximally entangled subsystem: $\ell>\mathcal{L}/N$}
When studying the dynamics from the crosscap state, the entanglement evolution for subsystems $\ell>\mathcal{L}/2$ could be obtained directly from that of $\ell<\mathcal{L}/2$ due to invariance of entanglement entropy under the swapping of the subsystem and its complement. For the multipodal entangled states, we need to consider separately the case of $\mathcal{L}/N<\ell<(N-1)\mathcal{L}/N$.  We start with the triangle state for which the correlation matrix is the same as in the previous subsection. It turns  out, however, that the calculation is much trickier since one must now account for the fact that the subsystem size scales with the total system size. In addition, we find that unlike the previous cases, the ratio of consecutive ${\rm tr}[\Gamma_N(t)^{2n}]$ does not admit a simple form. In particular, we find by inspecting the exact numerical results  ${\rm tr}[\Gamma_N(t)^{2n}]$ takes the form
\begin{eqnarray}\nonumber
    {\rm tr}[\Gamma_N(t)^{2n}]&=&\frac{1}{2^{2n}}\int_{-\pi}^{\pi} \frac{{\rm d}k}{2\pi}\big[  \chi_{L}(k,t)+ \chi_{2L}(k,t)+\chi_{-L}(k,t)+ \chi_{-2L}(k,t) \big]\\
    && +\Big(1-\frac{1}{2^{2n}}\Big)(\ell-L)+
    \Big(\frac{1}{2^{n-1}}-1\Big)\int_{-\pi}^{\pi} \frac{{\rm d}k}{\pi}\overline{\chi}(k,t)
\end{eqnarray}
wherein we recognize the same quasiparticle counting functions as in the previous section and we introduced 
\begin{eqnarray}
    \overline{\chi}(k,t)={\rm min}[2|v_k|t,\ell-L]~.
\end{eqnarray}
This new quasiparticle counting function is similar to those  appearing in quenches from lowly entangled states. In that context, it counts the number of quasiparticle pairs that are shared between a region and its complement. Here,  the counting function saturates at $\ell-L$ which is the size of the region within the subsystem which does not contribute to the initial entanglement. From the above expression, we see that, in common with the maximally entangled subsystem, the only $k$ dependence comes from the counting functions. Moreover,  the spectrum of $\Gamma_N(t)$ has a more complicated structure comprising several different eigenvalues not just $0,\pm 1/2$ as happened in the previous case. Indeed, initially 
$\Gamma_N(t)$  has $\ell-L$ eigenvalues $1$ with the rest being $0$ and as time passes some of the latter turn into $\pm 1/2$ through the first and second terms. The last term instead converts the eigenvalues $1$ into $\pm 1/\sqrt{2}$. 

\begin{figure}    
    \centering   
\centering   
    \centering   \includegraphics[width=0.49\linewidth]{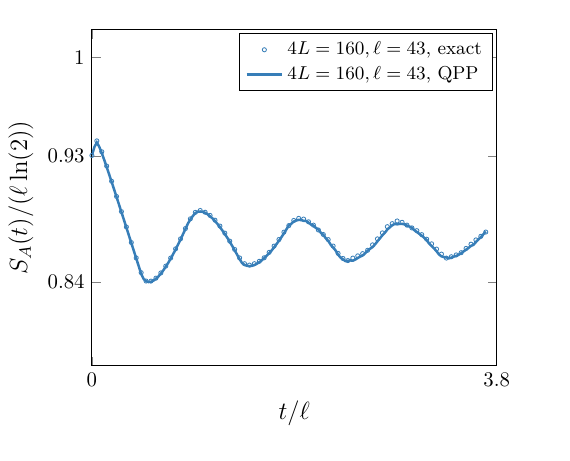}
    \includegraphics[width=0.49\linewidth]{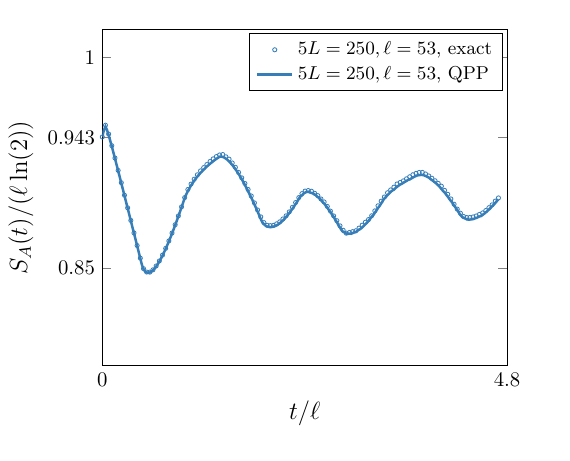}
       \includegraphics[width=0.49\linewidth]{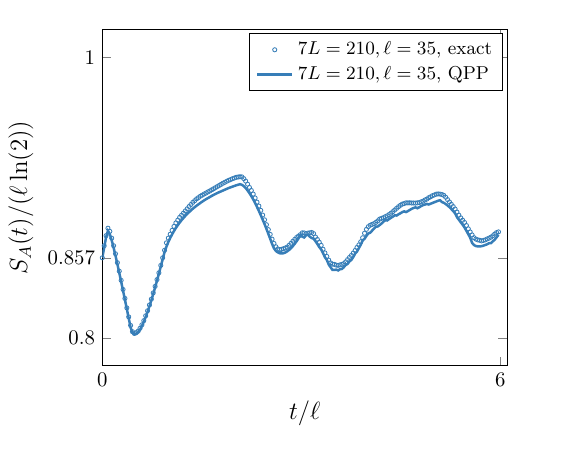}
       \includegraphics[width=0.49\linewidth]{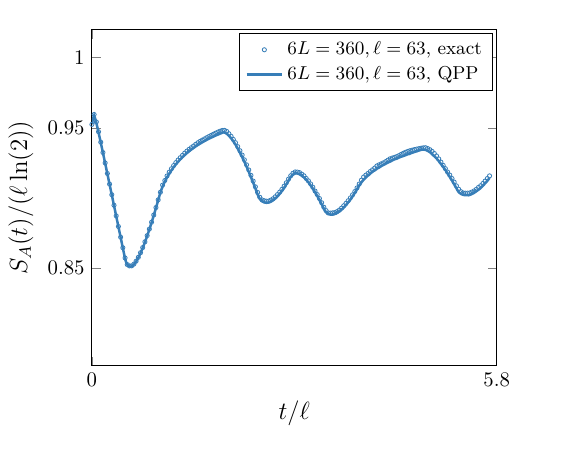}
            
    \caption{Entanglement dynamics for multipodal entangled states $N=4,5,6,7$ for $\ell>L$ clockwise from the top left. The symbols are the exact numerical results while the solid lines are the predictions of ~\eqref{eq:qpp_polygon_ell_non_max_ent}.   }
\label{fig:EE_single_subsys_polygonsbeta_ell_great_L}
\end{figure}

\begin{figure}    
    \centering   
\centering   
    \centering   
 \includegraphics[width=0.55\linewidth]{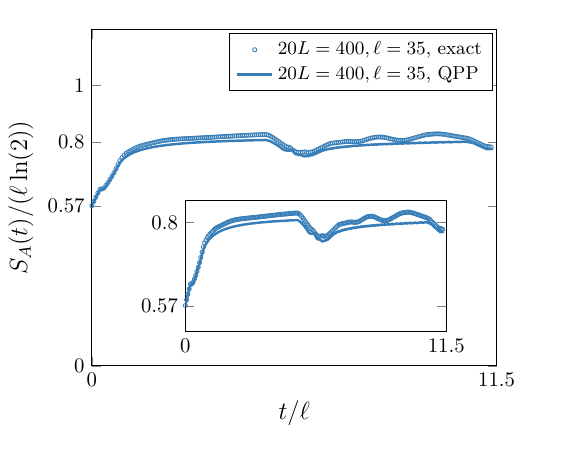}   
    
    \caption{Entanglement entropy dynamics of an ecosagon state $N=20$, with $L=20$. We see that since the subsystem $\ell>L$ the capacity it has to grow entanglement is significant and thus the entanglement entropy continues to increase in time, resulting in dynamics which resemble lowly entangled initial states. }
\label{fig:EE_ecosagon}
\end{figure}

After resumming the terms $\operatorname{tr}\left[\Gamma_N(t)^{2n}\right]$, the entanglement entropy is found to be
\begin{multline}
    S_A\big[\ket{\mathcal B_N(t)}\!\big]=\!L\log{2}+\frac{\ell-L}{2}\log{
    \!\left(\frac{27}{16}\right)}+ \Big[2\log{2}-\sqrt{2}\log{\left(1+\sqrt{2}\right)}\Big]\int\frac{{\rm d}k}{2\pi}\overline{\chi}(k,t)\\
    -\log{
    \!\left(\frac{27}{16}\right)\!}\!\int_{-\pi}^{\pi}
\frac{{\rm d}k}{8\pi}\!\left(\chi_{L}(k,t)+\chi_{\mathcal{L}-L}(k,t)+\chi_{-L}(k,t)+\chi_{L-\mathcal{L}}(k,t)\right)\,.
    \label{eq:qpp_polygon_ell_non_max_ent}
\end{multline}
We plot this expression for several different states in Figure~\ref{fig:EE_single_subsys_polygonsbeta_ell_great_L} and we compare it to the exact numerical results finding good agreement. We note that, unlike the maximally entangled subsystem which showed an initial period of constant entanglement, for $\ell>L$ the dynamics occurs immediately. Moreover, since the subsystem is not maximally entangled initially, it can increase with time rather than decrease. This initial increase arises from the new counting function which appears in the first line of~\eqref{eq:qpp_polygon_ell_non_max_ent} and can be accentuated by considering large $N$ as shown in Figure~\ref{fig:EE_ecosagon} where we show the dynamics for $N=20$. 

\begin{figure}    
    \centering   
\centering   
    \centering   \includegraphics[width=0.49\linewidth]{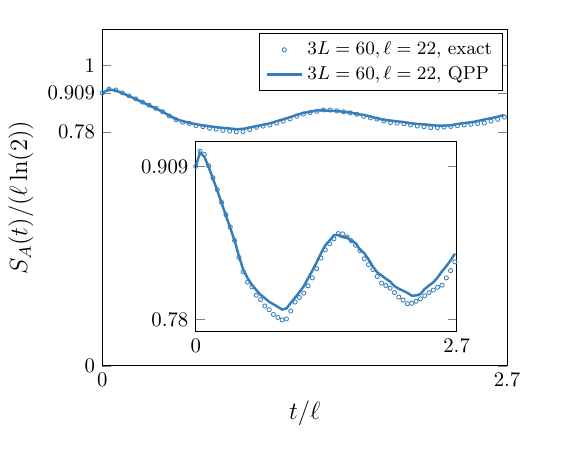}
    \includegraphics[width=0.49\linewidth]{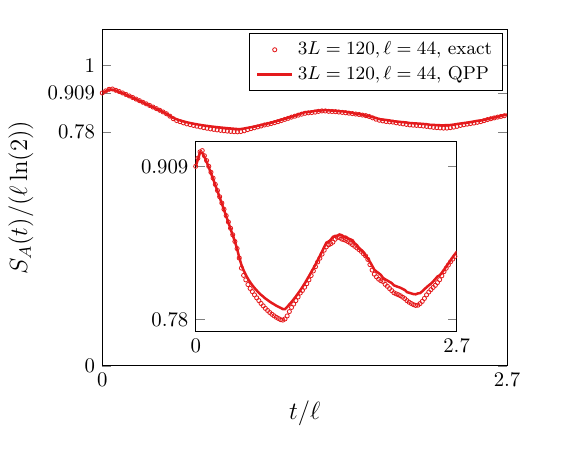}
    \includegraphics[width=0.49\linewidth]{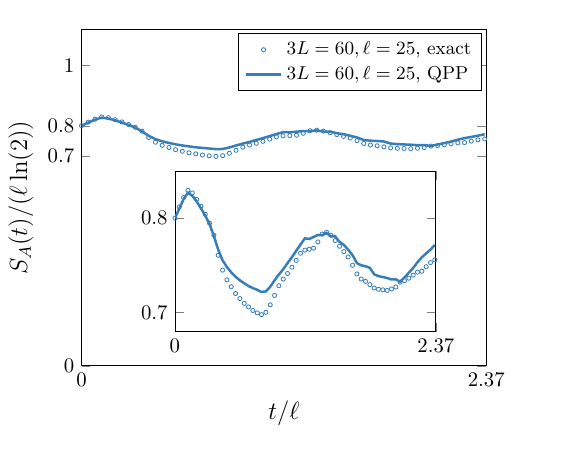}
    \includegraphics[width=0.49\linewidth]{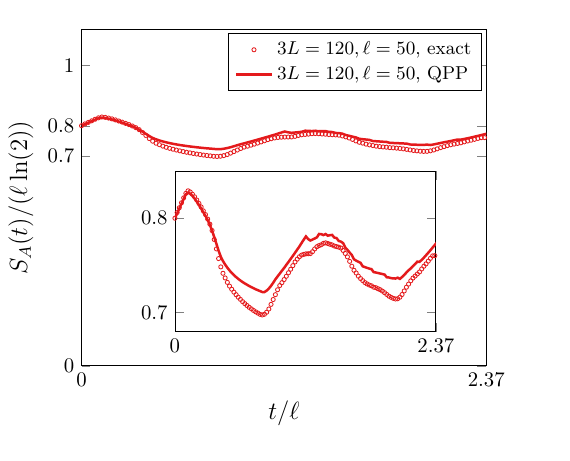}
    \caption{Triangle state dynamics for $\ell>L$ (non-maximally entangled subsystem initially), where we  see that the QPP prediction does not converge to the exact results for increasing subsystem and system size. Here we have the ratio $\delta_1=\frac{\ell-L}{\mathcal{L}}=0.0\bar{3}$. }
\label{fig:EE_single_subsys_delta1_triangle_ell_less_L}
\end{figure}

While the comparison between the numerics and the prediction~\eqref{eq:qpp_polygon_ell_non_max_ent} is quite good for some states and subsystem/system sizes, discrepancies can be seen. In Figure~\ref{fig:EE_single_subsys_delta1_triangle_ell_less_L} we plot the dynamics emerging from the triangle state, $N=3$ for several different values of $\ell,\mathcal{L}$. In the insets, we clearly see that some discrepancies are present which appear to remain finite upon doubling $\ell,\mathcal{L}$. Unlike quenches from lowly entangled states, for the multipodal entangled states we need to scale both time and subsystem size with $\mathcal{L}$ in order to see non-trivial dynamics. As a consequence, discriminating between finite size effects and deficiencies in the predictions~\eqref{eq:qpp_polygon} and ~\eqref{eq:qpp_polygon_ell_non_max_ent} can be difficult and requires further investigation. 


\begin{figure}    
    \centering   
\centering   
    \centering   \includegraphics[width=0.49\linewidth]{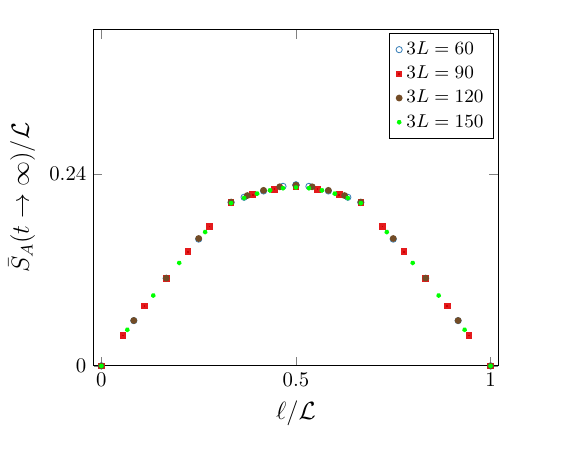}
\includegraphics[width=0.49\linewidth]{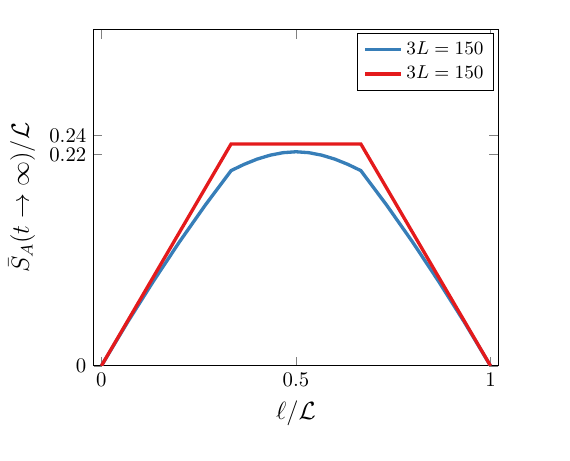} 
\includegraphics[width=0.49\linewidth]{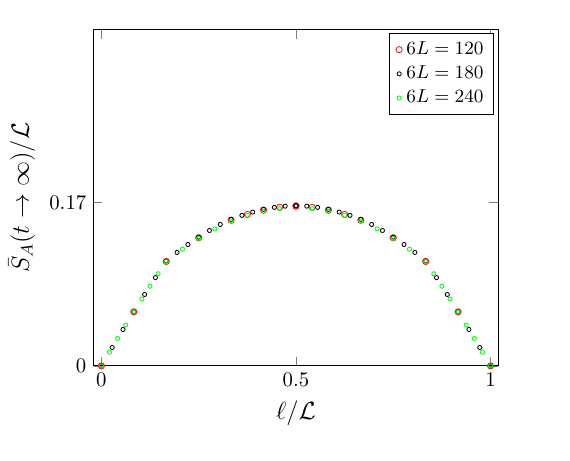}   \includegraphics[width=0.49\linewidth]{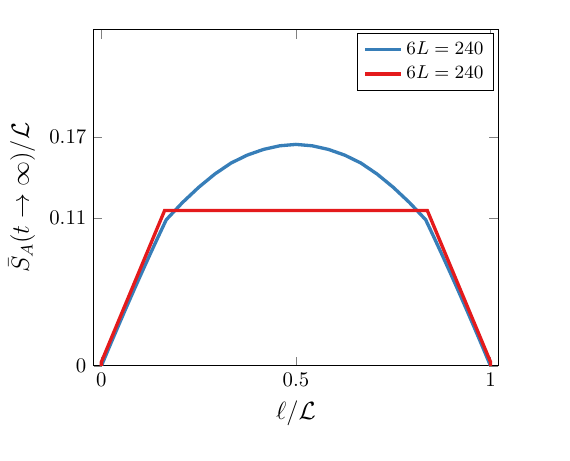}
   
    \caption{Time averaged value of entanglement entropy for the triangle (top) and hexagon states (bottom).
    Finite size scaling (left) shows that if properly scaled the graphs show collapse. On the right we compare to the initial value plotted in red.}
\label{fig:EE_time_avged1}
\end{figure}

\begin{figure}    
    \centering   
\centering   
    \centering   
 \includegraphics[width=0.6\linewidth]{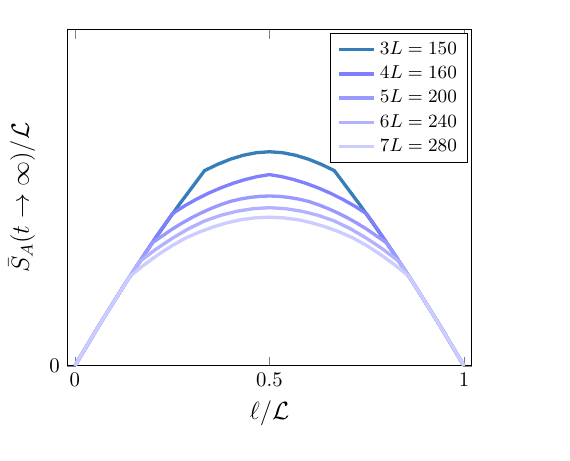}   
    
    \caption{Time averaged entanglement entropy for states with $N=3,4,5,6,7$ as a function of subsystem size, with both entanglement and subsystem size normalized by the size of the total system. We see that in the first regime of $\ell<L$ all states behave in the same manner, while for $\ell>L$ the curve is shifted for different $N$. }
\label{fig:EE_time_avged3}
\end{figure}

\textbf{Long time average:} The two slope structure of the initial state entanglement is felt in the long time average. It displays a distinct kink at $\ell=\mathcal{L}/N$ and another at $\ell=(1-1/N)\mathcal{L}$. 
In Figure~\ref{fig:EE_time_avged1} we plot the numerical long time average for different system sizes on the left for the triangle and hexagon states, $N=3,6$ . We see that upon rescaling all system sizes lie on the same line. On the right we plot a single set of numerical values compared to the initial state value. We see that for the triangle states the long time average lies below the initial value a feature which is not continued on to other $N$ as seen in the hexagon state. In Figure~\ref{fig:EE_time_avged3} we plot the long time average as a function of subsystem size for $N=3,4,5,6,7$. All curves lie on top of each other in the region $\ell<L$.

\subsection{Quasiparticle picture}
The entanglement dynamics of the crosscap state are transparently explained in terms of counter propagating entangled pairs of quasiparticles created at antipodal points. When such a pair is shared between $A$ and $\bar A$ they provide a $0$ eigenvalue to the correlation matrix $\Gamma_2(t)$ whereas if they are both contained within $A$ they contribute an eigenvalue of $1$. The value of $1$ indicates that the pair is not entangled with any other quasiparticles.  For the multipodal states the spectrum of $\Gamma_N(t)$ has a number of important differences. In particular, for a maximally entangled subsystem, the presence of the counting functions suggests that the dynamics can be captured by pairs of entangled quasiparticles. Importantly however, the eigenvalues of the correlation matrix do not switch from $0$ to $1$ but rather form $0$ to $1/2$. From this we infer that when the pair of quasiparticles are inside the subsystem they still retain some correlations and entanglement with $\bar A$. This is further evidenced by the fact that the coefficient of the time dependent terms in~\eqref{eq:qpp_polygon} is not $2\log(2)$ but $(1/4)\log(27/16)$. For subsystems which are not initially maximally entangled the subsystem exceeds the threshold which can support maximal entanglement and consequently $\Gamma_N(t)$ has a number of $1$ eigenvalues in the initial state equal $\ell-L$. At early times we see that the entanglement evolves due to the presence of the new counting function $\bar{\chi}$ which counts the number of quasiparticle pairs which are shared between the excess region and $\bar A$. Such pairs cause the initial $1$ eigenvalues to become $1/\sqrt{2}$ instead. 

From this we deduce that the dynamics of the system consists of multiplets of entangled quasiparticles which are created at the multipodal points dictated by the state $N$. We note that since the initial state is completely translationally invariant, the correlations can only exist between quasiparticles which have momenta $k$ and $-k$.  
This is in stark contrast to typical scenarios involving the presence of multiplets which require entanglement between momenta other than $k$ and $-k$.  For the multipodal states, the form of the counting functions supports the idea that only quasiparticles with momenta $k,-k$ are entangled. This behaviour is facilitated by the long range nature of the initial correlations which can cause the excitation of correlated quasiparticles with the same momenta but emerging from points separated by $L$. 
Note that since only two velocities are involved, $\pm v_k$, but we have more than two quasiparticles in our entangled multiplet, a number of entangled pairs within a multiplet will travel in the same direction and with the same velocity. Such pairs of quasiparticle chase each other around the system, remaining separated by their initial distance. Thus, their presence is felt not through a modification of the counting functions but through the fact that a pair inside the subsystem remains correlated with $\bar A$ at all times provided $\ell<L$. 

\begin{figure}    
    \centering   
\centering   
    \centering   
 \includegraphics[width=0.49\linewidth]{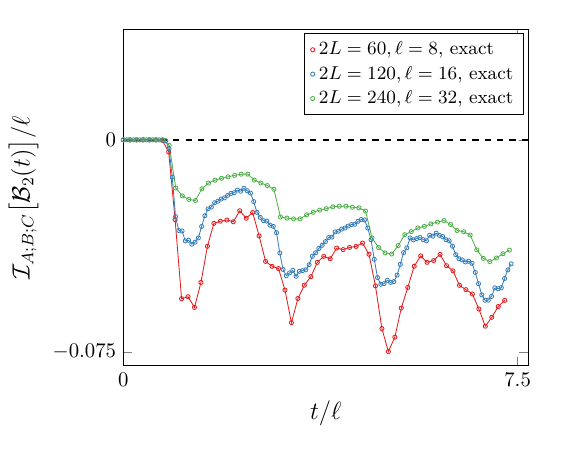}   
    \centering   
 \includegraphics[width=0.49\linewidth]{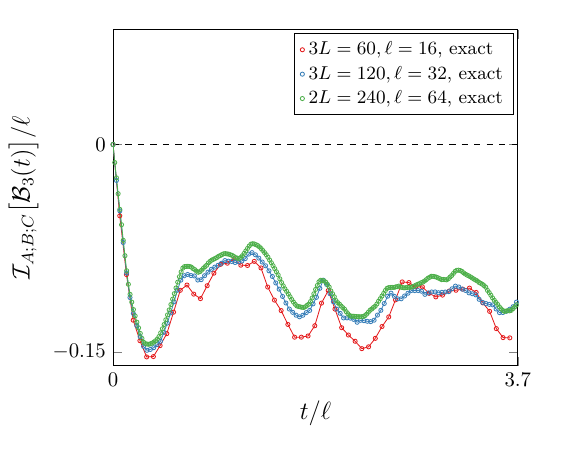}  
     \centering   
 \includegraphics[width=0.49\linewidth]{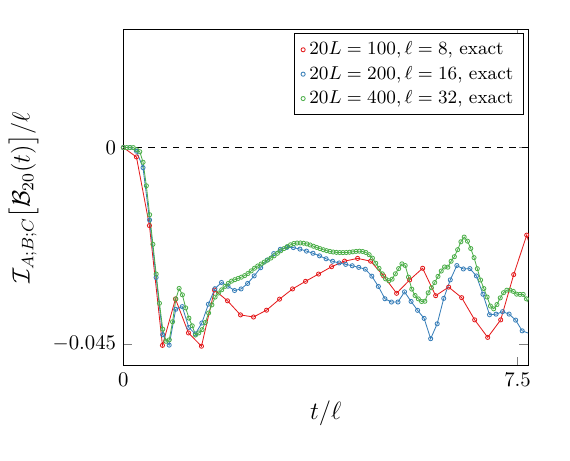}   
     \centering   
 \includegraphics[width=0.49\linewidth]{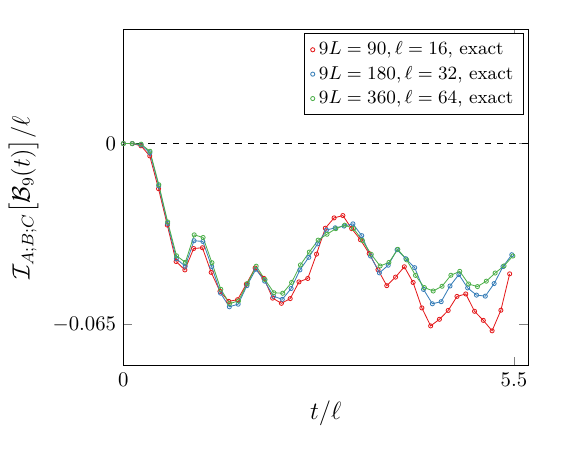}   
    \caption{The dynamics of the tripartite mutual information from different, entangled multipodal states and for different system and subsystem sizes. Clockwise from top left we depict the dynamics for the states with $N=2,3,9,20$ respectively. Here we see that for $N=2$ as $\ell$ and $L$ are increased $\mathcal{I}_{A;B;C}$ tends toward zero. For the multipodal states $N=3,9,20$ the same trend is not readily discernible. The symbols are exact numerics while the lines are a guide to the eye.  }
\label{fig:EE_mutual_info}
\end{figure}

\begin{figure}    
    \centering   
\centering   
    \centering   
 \includegraphics[width=0.85\linewidth]{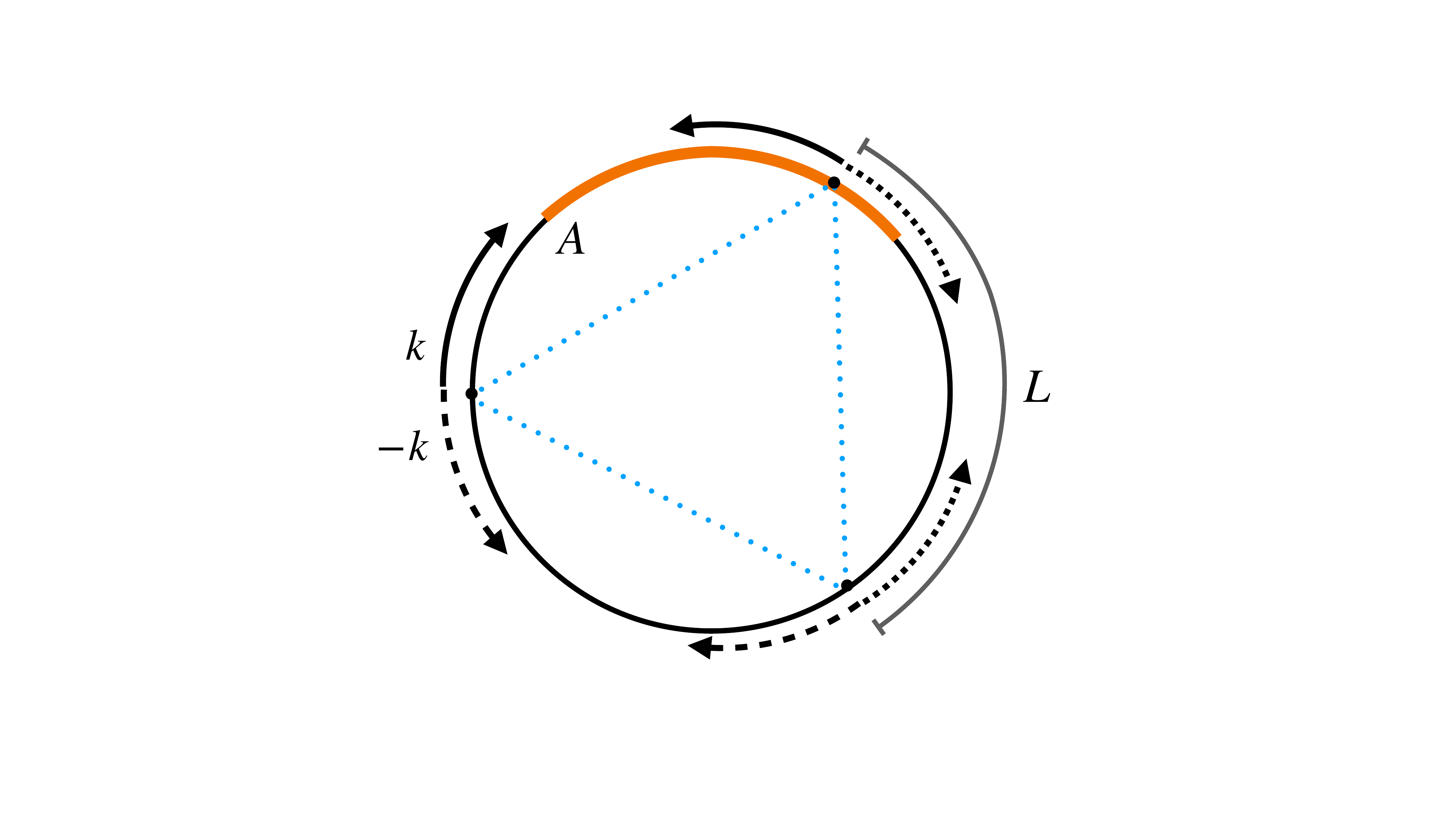}   
    
    \caption{A depiction of the quasiparticle dynamics of the entangled multipodal state. The quench excites $k,-k$ pairs (indicated by the arrows) throughout the chain but which have long range correlations between points which are $L$ sites apart. In the crosscap state only certain  pairings carry entanglement. These are depicted by the same type of arrow, solid, dotted or dashed. In the multipodal entangled states the negative tripartite information necessitates that there is entanglement between the pairs also.  }
\label{fig:QPP_multipode}
\end{figure}

Based on the above discussion, it is not evident how many quasiparticles make up the multiplet. To help with this, we examine the dynamics of the tripartite mutual information, $\mathcal{I}_{A;B;C}\big[\ket{\mathcal B_3(t)}\big]$ where $A,B,C$ are adjacent intervals. This quantity has been studied extensively in quenches to free fermion systems  from lowly entangled states~\cite{carollo2022entangled,caceffo2023negative,maric2023universality,maric2023universality2,berthiere2024tripartite,maric2023universality3}. Within the quasiparticle picture, it has been shown that this quantity requires that the quench produce multiplets of at least 4 quasiparticles to be non-zero in the ballistic scaling limit, i.e. for it to be extensive. Moreover, while a positive value is associated multiplets which are classically correlated, a negative value indicates that the quasiparticles are entangled~\cite{carollo2022entangled,caceffo2023negative}. We plot the exact numerics in Figure~\ref{fig:EE_mutual_info} for the crosscap state as well as the multipodal states with $N=3,9,20$ and for different subsystem and system sizes. Here we see that for all the states there is a  negative tripartite mutual information. In the crosscap state, as the system size is increased we observe the value decreases toward zero suggesting a sub-extensive tripartite  mutual information. This in agreement with the quasiparticle description of the crosscap quench in terms of entangled pairs. In contrast the same decreasing trend is not discernible in the $N=3,9,20$ states suggesting instead an extensive $\mathcal{I}_{A;B;C}[\ket{\mathcal{B}_{N>2}(t)}]$. A more complete analysis of the multipartite entanglement is required to make definitive claims, however the observed trends seem to support the existence of entangled multiplets of at least 4 quasiparticle in the multipodal states for $N>2$, see Figure~\ref{fig:QPP_multipode}.

\section{Conclusions}
In this work, we have introduced a class of Gaussian, translationally invariant states for periodic fermionic chains called multipodal entangled states. These states are generalizations of the crosscap or entangled antipodal pair states. While crosscap states feature correlations between antipodal points, the multipodal states exhibit correlations between multiple sites sitting at the vertices of an $N$-sided polygon. We study the entanglement properties of these states, both in equilibrium, and far from it after a quench to a free fermion model.

In equilibrium, we find that the multipodal states display a distinct entanglement profile as a function of subsystem size. For small subsystems $(\ell<\mathcal{L}/N)$ these states are maximally entangled. Above this value, the entanglement remains constant until we reach the reciprocal subsystem size, $\ell>(1-1/N)\mathcal{L}$, whereupon the entanglement decreases linearly to zero. In addition, the  multipodal states have non-maximal bipartite, mutual information, and zero tripartite information. All of this can be contrasted with crosscap states have maximal entanglement and bipartite mutual information, but zero tripartite information. The entanglement quench dynamics of multipodal states displays a richer structure in comparison to the crosscap states. In particular, two distinct regimes appear as function of the system size. For subsystems which are initially maximally entangled, the entanglement entropy remains constant for a certain period before dropping and then performing oscillations due to the finite size of the system. For larger subsystems, the entanglement evolves immediately and initially increases. 

The entanglement dynamics are very well captured by a quasiparticle expression which has some notable features. Although the state is single site translationally invariant and Gaussian, the quench produces multiplets of entangled quasiparticles  non-locally distributed throughout the system as opposed to entangled pairs as would be expected for lowly entangled initial states. Despite the multiplet structure, however,  the various dynamical processes are governed by the pairs of quasiparticles within the multiplet. The entanglement changes depending on whether a pair is fully inside a subsystem or not. Importantly, a pair which is fully inside the subsystem retains some entanglement with the rest of the multiplet which is outside the subsystem. Once again, this is a departure form the crosscap states whose dynamics is determined by pairs of maximally entangled quasiparticles. This picture is supported by time evolution of the tripartite mutual information which becomes negative indicating the presence of at least an entangled quadruplet structure. 

Our work invites numerous follow up studies.  These include investigating the symmetry properties of the multipodal states using the entanglement asymmetry~\cite{ares2023entanglement} or examining other measures of entanglement such as the entanglement negativity. Furthermore, one can also study some deformations of the multipodal states, akin to how the crosscap state is deformed to create a thermal pure state~\cite{yoneta_eap_PhysRevResearch.6.L042062,yoneta_eap_PhysRevLett.133.170404,bai2025spatially}. Lastly, in this work we have focussed on free fermion dynamics, however, it would be worthwhile to investigate other systems also, including CFTs, interacting integrable models or circuit dynamics.

\section*{Acknowledgments}
We thank Filiberto Ares, Bruno Bertini,  and Angelo Russotto for enlightening discussions.
PC and KC acknowledge support
by the ERC-AdG grant MOSE No. 101199196.

\appendix
\section{Momentum space squeezed state expression of $\ket{\mathcal{B}_{2}}$ and $\ket{\mathcal{B}_3}$}
\label{sec:appendix_crosscap_squeezed_state}
In this appendix, we go through the details of how to show that the crosscap and triangle crosscap states can be written as squeezed states.
This is crucial to realize how to generalize the states to general polygons with number of angles $N$ as discussed in the main text.
\subsection{The case of the crosscap state $\ket{\mathcal{B}_2}$}
We would like to see if we can express this state in momentum space as a squeezed state.
To this end, we first observe that using the anti-periodic boundary conditions $c^{\dagger}_{x+\mathcal{L}}=-c^{\dagger}_x$ we have the following equality
\begin{equation}
    \ket{\mathcal{B}_2}=\frac{1}{2^{L/2}}\Pi_{x=1}^{L}\left(1+c^{\dagger}_xc^{\dagger}_{x+L}\right)=\frac{1}{2^{L/2}}\Pi_{x=1}^{2L}\left(1+\frac{1}{2}c^{\dagger}_xc^{\dagger}_{x+L}\right),
\end{equation}
where the second state has the important difference that has a product going over $2L$ terms with anti-periodic conditions.

In this new expression of the state
we use the Fourier transform of the anti-periodic modes, given by
\begin{equation}
    c^{\dagger}_x=\frac{1}{\sqrt{2L}}\sum_{m=0}^{\mathcal{L}-1}c^{\dagger}_{k_m} e^{-ik_mx},\qquad k_m=\left(m+\frac{1}{2}\right)\frac{2\pi}{\mathcal{L}}
\end{equation}
and we have
\begin{eqnarray}
\ket{\mathcal{B}_2}&=&
    \frac{1}{2^{L/2}}\exp\left(\sum_{x,n_1,n_2=0}^{2L-1}\frac{1}{(2L)}\frac{1}{2}
    c_{k_{n1}}^{\dagger}c^{\dagger}_{k_{n2}}e^{i(k_{n1}+k_{n2})x+ik_{n2}L}
  \right)\ket{0} \\&=&\frac{1}{2^{L/2}}\exp\left(\sum_{n=0}^{2L-1}\frac{1}{2}
    c_{k_n}^{\dagger}c^{\dagger}_{2\pi-k_n}e^{-i(2\pi-k_n)L}
  \right)\ket{0}.
\end{eqnarray}

In order to write it as a squeezed state, we need to make sure to write the sum over positive modes $k_n>0$, thus we have
\begin{equation}
\ket{\mathcal{B}_2}=
\frac{1}{2^{L/2}}\exp\left(\sum_{n=0}^{2L-1}\frac{1}{2}
    c_{k_n}^{\dagger}c^{\dagger}_{2\pi-k_n}e^{-i(2\pi-k_n)L}
  \right)\ket{0}.
\end{equation}

Now, we need to organize the terms in the exponent in order for it to be a sum over positive momenta $k_n>0$.
In order to do so, first we split the sum over integers $n$ into two sums of equal length, leading to
\begin{eqnarray}
 \ket{\mathcal{\tilde{C}}}=
\frac{1}{2^{L/2}}\exp\left(\sum_{n=0}^{L-1}\frac{1}{2}
    c_{k_n}^{\dagger}c^{\dagger}_{2\pi-k_n}e^{ik_nL}
  + \sum_{n=L}^{2L-1}\frac{1}{2}
    c_{k_n}^{\dagger}c^{\dagger}_{2\pi-k_n}e^{ik_nL} \right)\ket{0}.   
\end{eqnarray}
Here, we focus on the second sum and redefine a new summation variable $m=n-L$ that also takes values from 0 to L-1,
\begin{eqnarray}
 \ket{\mathcal{\tilde{C}}}=
\frac{1}{2^{L/2}}\exp\left(\sum_{n=0}^{L-1}\frac{1}{2}
    c_{k_n}^{\dagger}c^{\dagger}_{2\pi-k_n}e^{ik_nL}
  + \sum_{m=0}^{L-1}\frac{1}{2}
    c_{k_m}^{\dagger}c^{\dagger}_{2\pi-k_m}e^{ik_mL} \right)\ket{0},   
\end{eqnarray}
where $k_m$ is as follows,
\begin{eqnarray}
    k_n=\left(n+\frac{1}{2}\right)\frac{2\pi}{3L}=\left(m+L+\frac{1}{2}\right)\frac{2\pi}{3L}=k_m+\pi\Rightarrow k_n=k_m+\pi.
\end{eqnarray}
By substitution we get
\begin{eqnarray}
 \ket{\mathcal{B}_2}=
\frac{1}{2^{L/2}}\exp\left(\sum_{n=0}^{L-1}\frac{1}{2}
    c_{k_n}^{\dagger}c^{\dagger}_{2\pi-k_n}e^{ik_nL}
  + \sum_{n=0}^{L-1}\frac{1}{2}
    c_{\pi+k_n}^{\dagger}c^{\dagger}_{\pi-k_n
    }e^{i(k_n+\pi)L} \right)\ket{0}.
\end{eqnarray}
Since we have a correspondence between $\pi-k_n$ and $k_n$, we move the two creation operators past each other, getting 
\begin{eqnarray}
 \ket{\mathcal{B}_2}=
\frac{1}{2^{L/2}}\exp\left(\sum_{n=0}^{L-1}\frac{1}{2}
    c_{k_n}^{\dagger}c^{\dagger}_{2\pi-k_n}e^{ik_nL}
  - \sum_{n=0}^{L-1}\frac{1}{2}
    c_{\pi-k_n}^{\dagger}c^{\dagger}_{\pi+k_n
    }e^{i(k_n+\pi)L} \right)\ket{0}.
\end{eqnarray}
For the term on the right we make the relabeling of 
\begin{eqnarray}
  \pi+ k_n\rightarrow2\pi-k_n\Rightarrow \pi-k_n\rightarrow k_n
\end{eqnarray}
getting
\begin{eqnarray}
 \ket{\mathcal{B}_2}=
\frac{1}{2^{L/2}}\exp\left(\sum_{n=0}^{L-1}\frac{1}{2}
    c_{k_n}^{\dagger}c^{\dagger}_{2\pi-k_n}e^{ik_nL}
  - \sum_{n=0}^{L-1}\frac{1}{2}
    c_{k_n}^{\dagger}c^{\dagger}_{2\pi-k_n
    }e^{i(2\pi-k_n)L} \right)\ket{0}, 
\end{eqnarray}
resulting in
\begin{eqnarray}
 \ket{\mathcal{B}_2}=
\frac{1}{2^{L/2}}\exp\left(\sum_{n=0}^{L-1}\frac{1}{2}
    c_{k_n}^{\dagger}c^{\dagger}_{2\pi-k_n}\left(e^{ik_nL}-e^{-ik_nL}\right)\right)\ket{0}.
\end{eqnarray}
Now, for anti-periodic boundary conditions imposed on the post-quench Hamiltonian, we work with modes that satisfy
\begin{eqnarray}
    k_n=\left(n+\frac{1}{2}\right)\frac{\pi}{L}\Rightarrow e^{ik_n2L}=-1\Rightarrow e^{-ik_nL}=e^{ik_nL}, 
\end{eqnarray}
so upon substitution we have
\begin{eqnarray}
 \ket{\mathcal{B}_2}=
\frac{1}{2^{L/2}}\exp\left(\sum_{n=0}^{L-1}e^{ik_nL}
    c_{k_n}^{\dagger}c^{\dagger}_{2\pi-k_n}\right)\ket{0}.
\end{eqnarray}
Here, we are in a position that we can write $\ket{\mathcal{B}_2}$ as a squeezed state
\begin{eqnarray}
   \ket{\mathcal{B}_2}=\frac{1}{2^{L/2}}\exp\left(\sum_{k>0}\mathcal{M}(k)
    c_{k}^{\dagger}c^{\dagger}_{2\pi-k}\right)\ket{0}=
\frac{1}{2^{L/2}}\exp\left(\sum_{n=0}^{L-1}e^{ik_nL}
    c_{k_n}^{\dagger}c^{\dagger}_{2\pi-k_n}\right)\ket{0},
\end{eqnarray}
with 
\begin{eqnarray}
    \mathcal{M}(k)=e^{ikL}\Rightarrow \mathcal{M}(-k)=-\mathcal{M}(k),
\end{eqnarray}
coming from the anti-periodic boundary conditions.

\subsection{The case of triangle crosscap state  $\ket{\mathcal{B}_3}$}

It is also important to understand the momentum space structure of $\ket{\mathcal{B}_3}$, which we rewrite here
\begin{equation}
    \ket{\mathcal{B}_3}=
    \frac{1}{4^{L/2}}\exp\left(\sum_{x=1}^{3L}\frac{1}{3}\left(
    c_{x}^{\dagger}c^{\dagger}_{x+L}+c_{x+L}^{\dagger}c^{\dagger}_{x+2L}+c_{x+L}^{\dagger}c^{\dagger}_{x+2L}\right)
  \right)\ket{0}.
\end{equation}
Our first aim is to show that we can bring $\ket{\mathcal{B}_3}$ in squeezed state form.
To do so, we perform a Fourier transform to write the state in Fourier space, using 
\begin{eqnarray}
    c^{\dagger}_{x}=\frac{1}{\sqrt{3L}}\sum_{n=0}^{3L-1}e^{-i k_nx}c^{\dagger}_{k_n},
\end{eqnarray}
leading to
\begin{multline}
\ket{\mathcal{B}_3}=\\
    \frac{1}{4^{L/2}}\exp\left(\frac{1}{3(3L)}\sum_{x,n_1,n_2=0}^{3L-1}
    c_{k_{n1}}^{\dagger}c^{\dagger}_{k_{n2}}e^{-i(k_{n1}+k_{n2})x}\left(e^{-ik_{n2}L} + e^{-2ik_{n2}L}- e^{-i(k_{n1}-k_{n2})L}  \right)\right)\ket{0}
\end{multline}
after using the antiperiodic conditions for $e^{-2ik_{n2}L}=-e^{ik_{n2}L}$.
Then, we use the Kronecker delta from the term on the outside,
\begin{eqnarray}
    \sum_{n_1,n_2=0}^{3L-1}
    e^{-i(k_{n1}+k_{n2})x}=3L\delta_{k_{n2},2\pi-k_{n1}}
\end{eqnarray}
leading to
\begin{multline}
\ket{\mathcal{B}_3}=\\
    \frac{1}{4^{L/2}}\exp\left(\frac{1}{3(3L)}\sum_{x,n_1=0}^{3L-1}
    c_{k_{n1}}^{\dagger}c^{\dagger}_{2\pi-k_{n1}}\left(e^{-i(2\pi-k_{n1})L} + e^{-2i(2\pi-k_{n1})L}- e^{-i(k_{n1}-2\pi+k_{n1})L}  \right)\right)\ket{0},
\end{multline}
which after simplifications and usage of antiperiodic conditions becomes
\begin{multline}
\ket{\mathcal{B}_3}=
    \frac{1}{4^{L/2}}\exp\left(\frac{1}{(3L)}\frac{1}{3}\sum_{x,n_1=0}^{3L-1}
    c_{k_{n1}}^{\dagger}c^{\dagger}_{2\pi-k_{n1}}\left(2e^{ik_{n1}L} + e^{2ik_{n1}L}  \right)\right)\ket{0}\\=
    \frac{1}{4^{L/2}}\exp\left(\frac{1}{(3L)}\sum_{x,n_1=0}^{3L-1}
    c_{k_{n1}}^{\dagger}c^{\dagger}_{2\pi-k_{n1}}\tilde{B}_3(k_{n1})\right)\ket{0},
\end{multline}
where it is easy to see that this form is not one of a squeezed state, since if we name 
\begin{eqnarray}
    \tilde{B}_3(k)=\frac{1}{3}\left(2e^{ik_{n1}L} + e^{2ik_{n1}L}\right),
\end{eqnarray}
does not satisfy the condition $\tilde{M}(-k)=-\tilde{M}(k)$.

In order to write it as a squeezed state, we need to make sure to write the sum over positive modes $k_n>0$, thus we have
\begin{equation}
\ket{\mathcal{\tilde{C}}}=
\frac{1}{2^{L/2}}\exp\left(\sum_{n=0}^{2L-1}\frac{1}{2}
    c_{k_n}^{\dagger}c^{\dagger}_{2\pi-k_n}e^{-i(2\pi-k_n)L}
  \right)\ket{0}.
\end{equation}
Following the same steps as before, by breaking the sum into two parts, rearranging the terms, and applying the anti-periodic boundary conditions, we get
\begin{eqnarray}
    \mathcal{B}_3(k)=e^{ikL}+e^{2ikL}.
\end{eqnarray}

\section{Generic squeezed states in momentum space}
\label{sec:appendix_squeezed_states}
Here we write some of the main results for fermionic squeezed states, which  are given by
\begin{equation}
    \ket{\Psi}=\mathcal{N}e^{\sum_{k>0}M(k)c^{\dagger}_{k}c^{\dagger}_{-k}}\ket{0},
\end{equation}
where $\ket{0}$ is the vacuum annihilated by all the operators $c_k$.

\subsection{Correlators in real and  momentum space for a squeezed state}
This conversion to a squeezed state in momentum space, leads to an efficient calculation of the correlation functions, in terms of the function $\mathcal{M}(k)$.
We already know that for a squeezed states the correlators are 
\begin{equation}
    \langle c^{\dagger}_kc_q\rangle=\frac{|\mathcal{M}(k)|^2}{1+|\mathcal{M}(k)|^2}\delta_{k,q}; \qquad  \langle c^{\dagger}_kc^{\dagger}_q\rangle=\frac{\mathcal{M}^{*}(k)}{1+|\mathcal{M}(k)|^2}\delta_{k,-q}, 
\end{equation}
which give us also the other two correlators
\begin{eqnarray}
       \langle c_kc^{\dagger}_q\rangle&=&\delta_{k,q}-\langle c^{\dagger}_qc_k \rangle=\delta_{k,q}-\frac{|\mathcal{M}(k)|^2}{1+|\mathcal{M}(k)|^2}\delta_{k,q}=\frac{1}{1+|\mathcal{M}(k)|^2}\delta_{k,q}; \\  \langle c_kc_q\rangle&=&\frac{\mathcal{M}(k)}{1+|\mathcal{M}(k)|^2}\delta_{k,-q}.
\end{eqnarray}

Then we can calculate the real-space correlators via the Fourier transform relations.
Doing all the correlation functions, we start from $\langle c^{\dagger}_{x} c_{y} \rangle$, given by
\begin{multline}
    \langle c^{\dagger}_{x} c_{y} \rangle=\sum_{m_{1,2}=0}^{\mathcal{L}-1}\langle c^{\dagger}_{k_{m1}}c_{k_{m2}}\rangle  \frac{e^{-ik_{m1}x+ik_{m2}y}}{2L} \\=\sum_{m_{1,2}=0}^{\mathcal{L}-1}\delta_{k_{m1},k_{m2}} \frac{e^{-ik_{m1}x+ik_{m2}y}}{2L}\frac{|\mathcal{M}(k_{m1})|^2}{1+|\mathcal{M}(k_{m1})|^2}=\sum_{m=0}^{\mathcal{L}-1} \frac{e^{-ik_{m}(x-y)}}{2L}\frac{|\mathcal{M}(k_{m})|^2}{1+|\mathcal{M}(k_{m})|^2},
\end{multline}
which also gives $\langle c_{x} c^{\dagger}_{y} \rangle$ by using the anti-commutation relations
\begin{multline}
    \langle c_{x} c^{\dagger}_{y} \rangle=\delta_{x,y}-\langle c^{\dagger}_{x} c_{y} \rangle=\sum_{m=0}^{\mathcal{L}-1}\frac{e^{-ik_m(x-y)}}{2L}-\sum_{m=0}^{\mathcal{L}-1} \frac{e^{-ik_{m}(x-y)}}{2L}\frac{|\mathcal{M}(k_{m})|^2}{1+|\mathcal{M}(k_{m})|^2}\\= \sum_{m=0}^{\mathcal{L}-1} \frac{e^{-ik_{m}(x-y)}}{2L}\frac{1}{1+|\mathcal{M}(k_{m})|^2}
\end{multline}
and then the anomalous correlations
\begin{multline}
    \langle c^{\dagger}_{x} c^{\dagger}_{y} \rangle=\sum_{m_{1,2}=0}^{\mathcal{L}-1}\langle c^{\dagger}_{k_{m1}}c^{\dagger}_{k_{m2}}\rangle  \frac{e^{-i(k_{m1}x+k_{m2}y)}}{2L} \\=\sum_{m_{1,2}=0}^{\mathcal{L}-1} \frac{\delta_{k_{m1},-k_{m2}}e^{-ik_{m1}x-ik_{m2}y}}{2L}\frac{\mathcal{M}^{*}(k_{m1})}{1+|\mathcal{M}(k_{m1})|^2}=\sum_{m=0}^{\mathcal{L}-1} \frac{e^{-ik_{m}(x-y)}}{2L}\frac{\mathcal{M}^{*}(k_{m})}{1+|\mathcal{M}(k_{m})|^2}, 
\end{multline}
which give the last correlator
\begin{equation}
    \langle c_xc_y\rangle=(\langle c^{\dagger}_yc^{\dagger}_x\rangle)^{\dagger}=\left[\sum_{m=0}^{\mathcal{L}-1} \frac{e^{-ik_{m}(y-x)}}{2L}\frac{\mathcal{M}^{*}(k_{m})}{1+|\mathcal{M}(k_{m})|^2}\right]^{*}=\sum_{m=0}^{\mathcal{L}-1} \frac{e^{-ik_{m}(x-y)}}{2L}\frac{\mathcal{M}(k_{m})}{1+|\mathcal{M}(k_{m})|^2}.
\end{equation}

The correlation matrix of a squeezed state is then given by
\begin{equation}
    C_{x,y}=(c^{\dagger}_x, c_x)^{T}(c_y,c^{\dagger}_y)=\left\langle\begin{pmatrix}
     c^{\dagger}_{x} c_{y} &  c_{x} c_{y}\\
c^{\dagger}_{x} c^{\dagger}_{y}  & c_{x} c^{\dagger}_{y}
     \end{pmatrix}\right \rangle
     =\frac{1}{2L}\sum_{m=0}^{\mathcal{L}-1} \frac{e^{-ik_m(x-y)}}{1+|\mathcal{M}(k)|^2} \begin{pmatrix}
   |\mathcal{M}(k)|^2  & \mathcal{M}(k)  \\
 \mathcal{M}^{*}(k)  & 1
 \end{pmatrix} .
\end{equation}

Another important form of the correlation matrix that is useful is $\Gamma_{x,y}$, which is defined as
\begin{equation}
    \Gamma_{x,y}=2C_{x,y}-\delta_{x,y},
\end{equation}
which for a squeezed state takes the form
\begin{equation}
   \Gamma_{x,y} = \frac{1}{2L}\sum_{m=0}^{\mathcal{L}-1} \frac{e^{-ik_m(x-y)}}{1+|\mathcal{M}(k)|^2} \begin{pmatrix}
   |\mathcal{M}(k)|^2-1  & 2\mathcal{M}(k)  \\
 2\mathcal{M}^{*}(k)  & 1- |\mathcal{M}(k)|^2
 \end{pmatrix}.
\end{equation}

\subsection{Time evolution of squeezed state with a free Hamiltonian}

In this case, we know that the evolution equation in the Heisenberg picture for the momentum space operators is
\begin{equation}
    c^{\dagger}_k(t)=e^{-i\epsilon(k)t}c^{\dagger}_{k}(0)
\end{equation}
and thus the time evolved correlation matrix is modified as follows
\begin{equation}
   \Gamma_{x,y}(t) = \frac{1}{2L}\sum_{m=0}^{\mathcal{L}-1} \frac{e^{-ik_m(x-y)}}{1+|\mathcal{M}(k)|^2} \begin{pmatrix}
   |\mathcal{M}(k)|^2-1  & 2\mathcal{M}(k)e^{-2i\epsilon(k)t}  \\
 2\mathcal{M}^{*}(k) e^{2i\epsilon(k)t} & 1- |\mathcal{M}(k)|^2
 \end{pmatrix},
\end{equation}
since only the anomalous correlators acquire a time evolution.

\section{Entanglement entropy dynamics}
\label{sec:appendix_entropy_dynamics_spa}
From the time evolved correlators we can compute the entanglement entropy, since we are interested in quenches of fermionic Gaussian states~\cite{peschel2003calculation}.
Here we include some computations using the method of the multi-dimensional stationary phase approximation of~\cite{fagotti2008evolution,calabrese2012quantum}.
The entanglement entropy between the subsystem $A$ and its complement $\bar{A}$, is computed through the reduced density matrix, which is \begin{equation}
    \rho_A(t)=\operatorname{Tr}_{\bar{A}}\left(\rho\right)
\end{equation}
and then $S_{A}(t)$ is the von-Neumann entanglement entropy given as,
\begin{equation}
    S_A(t)=-\operatorname{Tr}\rho_A\log{\rho_A}.
\end{equation}

For a Gaussian state, let us call it $\ket{\Psi}$, we know that the information about $\rho_{A}$ is encoded in the 2-point correlation matrix $C_{mn}$~\cite{peschel2003calculation}, which is a $2\mathcal{L}\times2\mathcal{L}$ matrix, defined as
\begin{equation}
    C_{mn}=\bra{\Psi}(c_m c^{\dagger}_m)(c^{\dagger}_n c_n)^{T}\ket{\Psi}.
\end{equation}
After we take its indices to be restricted to the subsystem $A$, we get $C_A$, which is a $2\ell\times2\ell$ matrix, that in the case of pure states, has $2\ell$ eigenvalues $\lambda_n$ in the interval $\lambda_n\in[0,1]$.
Numerically, the entanglement entropy can be calculated by diagonalizing the 2-point correlation matrix, using the following relation
\begin{equation}
    S_{A}(t)=-\sum_n\left[\lambda_{n}\log{\lambda_n}+(1-\lambda_n)\log{(1-\lambda_n)}\right].
\end{equation}
There is another approach to calculate it, which gives us analytical control over the result, again based on the correlation matrix.
First, we define $\Gamma_{mn}$ as
\begin{equation}
    \Gamma_{mn}=2C_{mn}-\mathbf{1},
\end{equation}
where its eigenvalues $\nu_n=2\lambda_n-1\in[-1,1]$.
When we take the restriction of its indices to be witihn the subsystel $A$, which we call $\Gamma_A$,  we can write the entanglement entropy as
\begin{equation}
    S_{A}(t)=-\sum_n\left[\frac{1+\nu_{n}}{2}\log{\left(\frac{1+\nu_{n}}{2}\right)}+\frac{1-\nu_{n}}{2}\log{\left(\frac{1-\nu_{n}}{2}\right)}\right]
\end{equation}
In terms of the full matrix $\Gamma_A$, the entanglement entropy can be written as
\begin{equation}
    S_{A}(t)=-\operatorname{Tr}\left[\frac{1+\Gamma_A}{2}\log{\left(\frac{1+\Gamma_A}{2}\right)}+\frac{1-\Gamma_A}{2}\log{\left(\frac{1-\Gamma_A}{2}\right)}\right],
\end{equation}
which is the starting point of the calculation.

\subsection{Expanding the logarithms }
Starting from the equation
\begin{equation}
    S_{A}(t)=-\operatorname{Tr}\left[\frac{1+\Gamma_A}{2}\log{\left(\frac{1+\Gamma_A}{2}\right)}+\frac{1-\Gamma_A}{2}\log{\left(\frac{1-\Gamma_A}{2}\right)}\right],
\end{equation}
we can expand the logarithms in powers of the matrix $\Gamma_A$ as
\begin{equation}
   \
   \log{\left(1+\Gamma_A\right)}=\sum_{k=1}^{\infty}\frac{(-1)^{k}}{k}\Gamma_A^{k}, \qquad \log{\left(1-\Gamma_A\right)=-\sum_{k=1}^{\infty}\frac{\Gamma_A^{k}}{k}}
\end{equation}
giving
\begin{equation}
     S_{A}(t)=-\operatorname{Tr}_{A}\left[ -\sum_{k=1}^{\infty}\frac{\Gamma_A^{2k}}{2k(2k-1)}-\log{2} \right]   = \sum_{k=1}^{\infty}\frac{\operatorname{Tr}_{A}\left[ \Gamma_A^{2k} \right]}{2k(2k-1)}+\operatorname{Tr}_{A}\left[\log{2}\right], 
\end{equation}
this shows us that the information of entanglement entropy is encoded in the trace of the even powers of the matrix $\Gamma_A$.
So, these are the objects that we should calculate at this point.

For Gaussian states, we can expand in traces of powers of the correlation matrix, since we have the relation
\begin{equation}
    S_{A}(t)=-\frac{1}{2}\operatorname{Tr}\left[\frac{1+\Gamma_A}{2}\log{\frac{1+\Gamma_A}{2}}+\frac{1-\Gamma_A}{2}\log{\frac{1-\Gamma_A}{2}}\right],
\end{equation}
by expanding in powers of $\Gamma_A$, we have
\begin{equation}
    S_{A}(t)=\ell\log{2}-\frac{1}{2}\sum_{n=1}^{\infty}\frac{1}{2n(2n-1)}\operatorname{Tr}\left[\Gamma_A^{2n}(t)\right].
\end{equation}
This is the relation on which the stationary phase approximation is based.

\bibliographystyle{ytphys}
\bibliography{mybib}

\end{document}